\newcommand{\greatTitle}{A Comprehensive Workflow for General-Purpose Neural Modeling with Highly Configurable Neuromorphic Hardware Systems}

\documentclass[twocolumn]{svjour3}
\usepackage{mathptmx} %
\journalname{}

\usepackage{psfrag}
\usepackage{epsfig}
\usepackage{graphicx}
\usepackage{url}
\usepackage[round,colon]{natbib}
\usepackage{listings}
\usepackage{subfig}
\usepackage{units}
\usepackage{booktabs}
\usepackage[utf8]{inputenc}
\usepackage[T1]{fontenc}
\usepackage[ps2pdf,bookmarks]{hyperref}
\hypersetup{
      pdftitle={\greatTitle},
      pdfsubject={},
      pdfauthor={Br\"uderle et al.},
      pdfkeywords={},
      pdfcreator={},
      pdfproducer={LaTeX},
      colorlinks=true,
      urlcolor=blue,
      linkcolor=black,
      citecolor=black,
      urlcolor=black,
      breaklinks=true
}
\usepackage{breakurl}
\usepackage{upgreek}
\usepackage{float}
\usepackage{paralist}
\usepackage{mdwlist}

\usepackage{stfloats}
\usepackage{fixltx2e}
\usepackage{afterpage}

\usepackage{color}

\usepackage[]{todonotes} %

\usepackage{lipsum}
\setlipsumdefault{1}
\newcommand{\CPP}{C\kern-0.05em\raisebox{0.10em}{\scalebox{0.8}{++}}}

\title{\greatTitle}

\newcommand{\AllAuthorNames}{
Daniel Br\"uderle \and
Mihai A. Petrovici \and
Bernhard Vogginger \and
Matthias Ehrlich \and
Thomas Pfeil \and
Sebastian Millner \and
Andreas Gr\"ubl \and
Karsten Wendt \and
Eric M\"uller \and
Marc-Olivier Schwartz \and
Dan Husmann de Oliveira \and
Sebastian Jeltsch \and
Johannes Fieres \and
Moritz Schilling \and
Paul M\"uller \and
Oliver Breitwieser \and
Venelin Petkov \and
Lyle Muller \and
Andrew P.~Davison \and
Pradeep Krishnamurthy \and
Jens Kremkow \and
Mikael Lundqvist \and
Eilif Muller \and
Johannes Partzsch \and
Stefan Scholze \and
Lukas Z\"uhl \and
Christian Mayr \and
Alain Destexhe \and
Markus Diesmann \and
Tobias C.~Potjans \and
Anders Lansner \and
Ren\'e Sch\"uffny \and
Johannes Schemmel \and
Karlheinz Meier
}

\newcommand{\AllAuthorAffiliations}{
D.~Br\"uderle \and O.~Breitwieser \and J.~Fieres \and A.~Gr\"ubl \and D.~Husmann de Oliveira \and S.~Jeltsch \and K.~Meier \and S.~Millner \and E.~M\"uller \and P.~M\"uller \and V.~Petkov \and M.~A.~Petrovici \and T.~Pfeil \and J.~Schemmel \and M.~Schilling \and M.~Schwartz \and B.~Vogginger \at Kirchhoff Institute for Physics\\Ruprecht-Karls-Universit\"at Heidelberg, Germany\\
Tel.: +49 6221 549813\\
\email{bruederle@kip.uni-heidelberg.de}
\and
\emph{Present address} of M.~Schilling \at Robotics Innovation Center, DFKI Bremen, Germany
\and
A.~P.~Davison \and A.~Destexhe \and L.~Muller \at Unit\'e de Neuroscience, Information et Complexit\'e, CNRS, Gif sur Yvette, France
\and
M.~Diesmann \at RIKEN Brain Science Institute and RIKEN Computational Science Research Program, Wako-shi, Japan\\Bernstein Center for Computational Neuroscience, Universit\"at Freiburg, Germany
\and
M.~Ehrlich \and C.~Mayr \and J.~Partzsch \and S.~Scholze \and R.~Sch\"uffny \and K.~Wendt \and L.~Z\"uhl \at Institute of Circuits and Systems, Technische Universit\"at Dresden, Germany
\and
J.~Kremkow \at Bernstein Center Freiburg, University of Freiburg, Germany
\and
P.~Krishnamurthy \and A.~Lansner \and M.~Lundqvist \at Computational Biology, KTH Stockholm, Sweden
\and
E.~Muller \at Brain Mind Institute, Ecoles Polytechniques Federales de Lausanne, Switzerland
\and
T.~C.~Potjans \at Institute of Neuroscience and Medicine (INM-6), Research Center J\"ulich, Germany\\RIKEN Computational Science Research Program, Wako-shi, Japan
}

\begin{document}

\title{
\greatTitle
}
\author{
	\AllAuthorNames
}

\institute{
	\AllAuthorAffiliations
}

\date{}

\maketitle

\begin{abstract}
In this paper we present a methodological framework that meets novel requirements emerging from upcoming types of accelerated and highly configurable neuromorphic hardware systems.
We describe in detail a device with 45 million programmable and dynamic synapses that is currently under development, and we sketch the conceptual challenges that arise from taking this platform into operation.
More specifically, we aim at the establishment of this neuromorphic system as a flexible and neuroscientifically valuable modeling tool that can be used by non-hardware-experts.
We consider various functional aspects to be crucial for this purpose, and we introduce a consistent workflow with detailed descriptions of all involved modules that implement the suggested steps: 
The integration of the hardware interface into the simulator-independent model description language PyNN; 
a fully automated translation between the PyNN domain and appropriate hardware configurations; 
an executable specification of the future neuromorphic system that can be seamlessly integrated into this biology-to-hardware mapping process as a test bench for all software layers and possible hardware design modifications; 
an evaluation scheme that deploys models from a dedicated benchmark library, compares the results generated by virtual or prototype hardware devices with reference software simulations and analyzes the differences. 
The integration of these components into one hardware-software workflow provides an ecosystem for ongoing preparative studies that support the hardware design process and represents the basis for the maturity of the model-to-hardware mapping software.
The functionality and flexibility of the latter is proven with a variety of experimental results.
\keywords{Neuromorphic \and VLSI \and Hardware \and Wafer-Scale \and Software \and Modeling \and Computational Neuroscience \and PyNN}
\end{abstract}

\section{Introduction}
\label{section:introduction}

\subsubsection*{Neuroscience and Technology} 
Advances in neuroscience have often gone hand in hand with significant progress in the applied technologies, tools and methods.
While the experimental investigation of living neural tissue is indispensable for the generation of a detailed knowledge base of the brain, from which understanding of underlying principles can emerge, technological difficulties have always imposed limits to this endeavor. 
Until today it is not possible to study relevant observables in a sufficiently large fraction of brain tissue under realistic conditions and with a spatiotemporal resolution that is high enough to fully capture -- and possibly consistently explain -- the mechanisms of higher order brain functions.

Therefore, in neuroscience, like in any other research field on dynamical systems that cannot be fully explored by experimental methods, models represent an indispensable approach to test hypotheses and theories on the real subject of interest.
However, even neural modeling is significantly constrained and influenced by the set of available technologies.
The spectrum of feasible experimental setups, in particular in \textit{computational neuroscience}, 
directly depends on the accessible computational power.
The difficulty of efficiently mapping the massive parallelism of neural computation in biological tissue to a limited number of digital \textit{general purpose} CPUs is a crucial bottleneck in the development of large-scale computational models of neural networks, where statistics-intensive analyses or long-term observations of network dynamics can become computationally extremely expensive (see e.g.\ \citealp{morrison05distributed, brette06simulators, morrison07stdp}).

\subsubsection*{Neuromorphic Hardware} 
For an alternative modeling approach, the so-called \textit{neuromorphic engineering}, the technology-driven nature is even more obvious.
In a physical, typically silicon form, neuromorphic devices mimic the structure and emulate the function of biological neural networks. 
This branch of neuroscience has its origins in the 1980s \citep{mead88silicon, mead89analog, mead90neuromorphic}, and today an active community is working on analog or mixed-signal \textit{VLSI}\footnote{Very Large Scale Integration} models of neural systems (for reviews see e.g.\ \citealp{renaud2007neuromimetic, indiveri2009artificial}).

Dedicated implementations of said computational models are typically more power efficient compared to general purpose architectures and are well suited for e.g.\ embedded controllers of autonomous units like robots.
Fault tolerance features observed in biological neural architectures are expected to apply to corresponding neuromorphic hardware implementations as well.
This fact can offer one important way to create reliable computing components on the basis of future nano-scale hardware constituents, where current design strategies will run into serious yield problems.
Moreover, the inherent parallelism of on-chip emulation of neural dynamics has the potential to overcome the aforementioned scaling limitations of pure software simulations. 

Still, until today the focus of neuromorphic projects is mostly very application-specific. 
The majority of groups is working on neuromorphic sensors like e.g.\ silicon retinas and visual processing systems \citep{netter02arobotic, delbrueck04silicon, serrano_nips2005, merolla2006dynamic, zhengming08, gomez2010realtime} or motor control in robotics \citep{lewis00toward}. 
The requirement of communication with the environment is one important reason for the fact that nearly all neuromorphic devices reported so far are designed to operate in real-time.
But even the projects that deal with mimicking, studying or applying neural information processing \citep{vogelstein2007reconfigurable}, self-organization \citep{hafliger2007adaptive, mitra09realtime} or even hybrid setups coupling neuromorphic devices with living tissue \citep{bontorin2007realtime} are usually focused on one type of neural architecture, one anatomical region or one function the implemented network is supposed to fulfill.

Two main reasons for this self-limitation of neuromorphic development are the finite size of every neuromorphic device as well as the limited possibilities to change the behavior of individual cells and the network connection patterns once they have been cast into silicon.
A typical approach to reduce size limitations is to scale up networks by inter-connecting multiple hardware modules \citep{serrano_tcas2005, berge_iscas07, indiveri2008_vlsi, schemmel_ijcnn2008}.
Furthermore, recent advances in neuromorphic development eventually promise to overcome the limited flexibility of hardware models by offering a sufficiently fine-grained configurability of both the neuron parameter values as well as the network connectivity \citep{indiveri_tnn2006, schemmel_iscas07, ehrlich_ssd07, schemmel_ijcnn2008, indiveri2009artificial, schemmel_iscas2010}.
This crucial feature allows to consider the utilization of neuromorphic systems as flexible modeling tools to approach open neuroscientific questions with new strategies \citep{kaplan_ijcnn2009, bruederle09establishing_hack, bruederle09phd, bruederle10simulator}.

\subsubsection*{A Novel Methodological Approach}
The FACETS\footnote{Fast Analog Computing with Emergent Transient States} research project \citep{facetsproject_homepage} and its successor BrainScaleS \citep{brainscales_homepage} aim at a comprehensive exploitation of the possibilities inherent to that approach.
The highly interdisciplinary collaborations gather neurophysiological, theoretical and hardware expertise in order to develop and operate a large-scale neuromorphic device that can serve as a flexible neural network emulation platform with hitherto unattained configurability and acceleration.
It is planned to exploit this combination of features with experimental paradigms that are not realizable with pure software simulations, like long-term learning studies, systematic parameter explorations and the acquisition of statistics for every tested setup.

Following this attempt, one important insight has emerged that has only rarely been addressed in the literature so far (exceptions are e.g.\ \citealp{dante2005hardsoft, oster2005hardsoft}): 
Any hardware device that is complex enough to serve as a useful neural modeling tool is useless without an appropriate software environment that implements a reasonable methodological framework for its operation. 
For any developed neuromorphic modeling platform, hard- and software have to form a functional unit.
Moreover, the need for methods that have to be applied in order to make the advantages of a neuromorphic device accessible to non-hardware experts does not only refer to the actual \emph{operation} of the device itself.
Instead, already its \emph{design process} needs to be supported and influenced by preparatory studies, e.g.\ with virtual versions of the future hardware-software system.

In this publication we summarize the FACETS efforts to create a comprehensive methodological framework providing a workflow aiming to make the innovative FACETS wafer-scale hardware system a generic modeling tool that is applicable to a wide range of neuroscientific questions and accessible to the neuroscientific community.

\subsubsection*{Structure of this Article}
This introduction is followed by a description of the complete neuromorphic modeling infrastructure.
This includes both the utilized hardware devices and the \emph{workflow} that is in focus of this paper, i.e.\ the framework of methods and software modules that have been developed for the design-assistance, the benchmarking and the actual operation of the platform.
A third section presents data and results that provide a proof of functionality for the concept as a whole.
Various components of the workflow are evaluated, and the performance of benchmark model experiments performed with the complete system is studied and analyzed.
The last section discusses the state of validation of the presented framework as well as its advantages and limitations considering alternative approaches.
Implications and plans for future work and new perspectives arising from the presented achievements are outlined.

\section{The Workflow Components: Modules and Methods}
\label{section:methods}

\begin{figure*}[tbp]
    \centering
    \includegraphics[width=.92\linewidth]{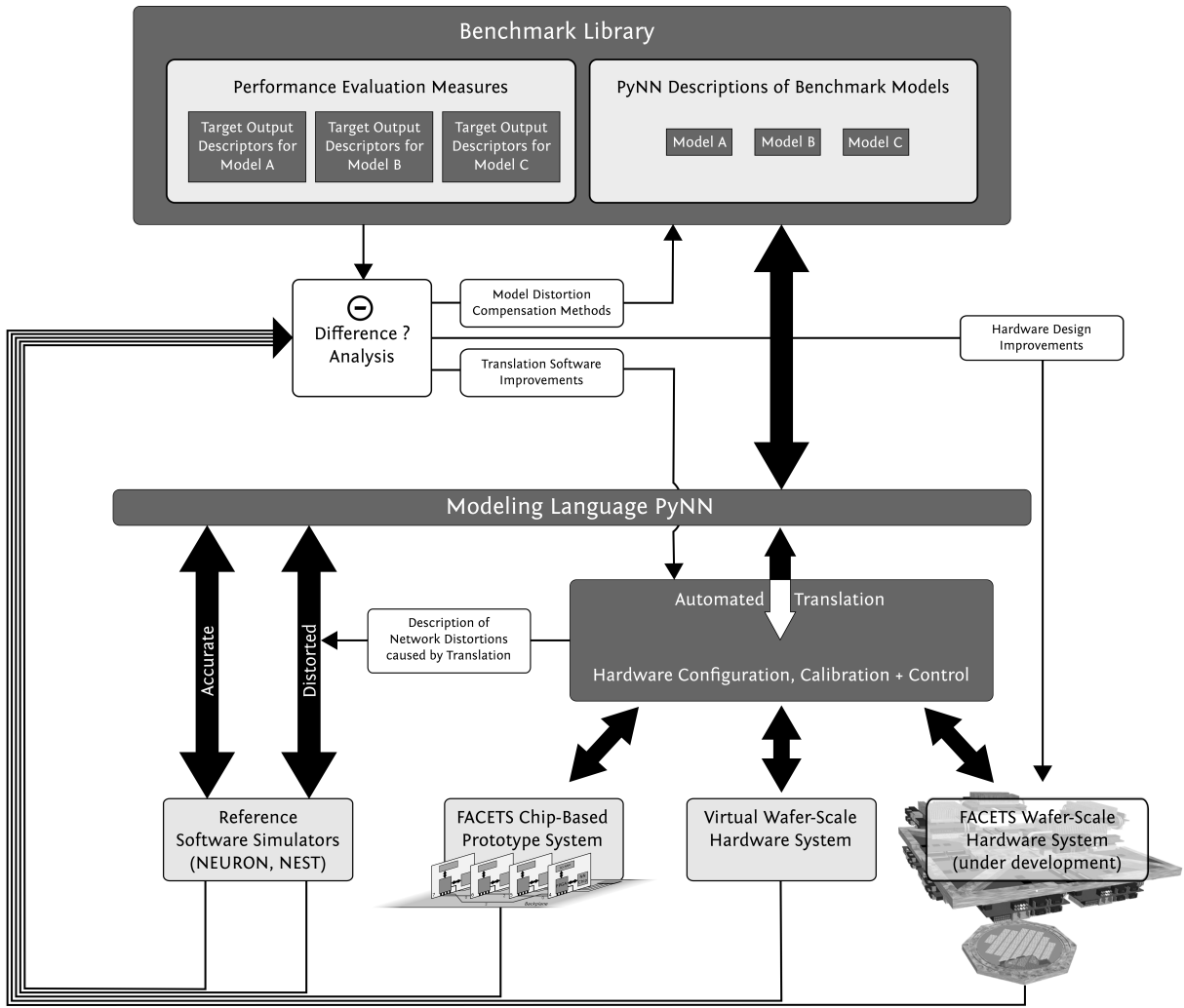}
    \caption{\label{figure:methods_overview} Optimization workflow towards neural modeling with neuromorphic hardware devices. The main components are 1.~the highly configurable FACETS neuromorphic hardware devices, 2.~the software module stack that performs an automated translation of neural network experiments described with the modeling language PyNN into corresponding hardware configuration and control patterns, and 3.~a benchmark library that contains a collection of neuroscientific models written in PyNN. For a detailed explanation of the complete flow and all individual steps and components see full text.}
\end{figure*}

The following section provides an overview over the complete infrastructure that has been developed to realize a novel neural modeling concept built around the FACETS neuromorphic hardware system.
For this purpose, the neuromorphic device itself is presented in Section~\ref{section:hardware} on a level of detail that is appropriate to the method descriptions that follow. 
These methods are either implemented by or directly rely on an innovative software framework, which will be explained in Section~\ref{section:software} by means of its structure and concepts.
A significant achievement for the targeted design and development of a harmonizing hardware-software unit forming the modeling platform was the collection of a set of literature-based benchmark model experiments, as summarized in Section~\ref{section:benchmarks}.

The workflow that has been developed around these three main components is schematically depicted in Figure~\ref{figure:methods_overview}:
The library of dedicated neuroscientific benchmark models, including descriptions and measures to evaluate their correct functionality, has been established by members of the FACETS research project \citep{facetsproject_homepage}. 
For any model from within this set, a description on the basis of the simulator-independent modeling language PyNN (see Section~\ref{section:pynn}) is available. 
The mentioned translation software stack performs an automated conversion of these scripts into appropriate data for the configuration and control of different hardware or software back-ends. 
The same stack also re-translates the resulting hardware output into the domain of its biological interpretation. 
During the development and optimization phase of the FACETS wafer-scale hardware system, an elaborate virtual version of the device (see Section~\ref{section:virtual_hardware}) serves as a test bench for the development and tuning of all involved translation software modules. 

In addition to the virtual wafer-scale device, a purely chip-based neuromorphic system (see Section~\ref{section:stage1hardware}) provides important information about characteristics of circuits planned to be implemented in the wafer-scale system.
These ASICs thereby support the wafer-scale design process and the development of possible strategies to compensate unavoidable phenomena like transistor-level variations or noise. 
The outputs of all applied hardware or virtual hardware back-ends are compared with the target output descriptions included with the models in the benchmark library and with reference experiments on pure software simulators. 
The remaining differences are analyzed, as is exemplarily presented in Section~\ref{section:benchmark_results}.

In an ongoing optimization flow, the benchmark models are repeatedly mapped to the still evolving hardware substrate with the likewise continuously optimized software framework.
The iteratively applied result analyses provide the fundament for improvements that close the workflow loop: The hardware design, the biology-to-hardware translation modules and optionally even the models themselves are modified such that the functional effects of remaining distortions caused by the model-to-hardware mapping process are minimized.

Hence, the first application of the presented workflow is to take novel types of hardware devices into operation.
Furthermore, it can serve as a basic methodological paradigm for the actual target application of neuromorphic systems, i.e.\ the exploration and optimization of neural architectures by means of different optimization objectives.
These include the search for computationally powerful structures or for setups that can reproduce biologically plausible dynamics.

While this section motivates and explains the workflow \emph{as a whole} and provides descriptions of all involved components and methods,
the scope of this paper would be exceeded by providing detailed motivation for \emph{all} particular choices of methods and components being part of the framework.
The reasons for individual methodological or design decisions can be found in the literature referenced within the corresponding paragraphs.

\subsection{The FACETS Hardware System}
\label{section:hardware}

In the following, the FACETS wafer-scale hardware system will be described with focus on conceptual and technical details that are relevant in the context of this article.
More information on the hardware setup and circuitry can be found in \cite{schemmel_ijcnn2008}, \cite{ehrlich_ssd07}, \cite{millner10} and \cite{schemmel_iscas2010}.

At the core of the FACETS wafer-scale hardware system (see Figure~\ref{figure:waferscale_system}) is an uncut wafer built from mixed-signal ASICs\footnote{Application Specific Integrated Circuit}, named \textit{High} \textit{Input} \textit{Count} \textit{Analog} \textit{Neural} \textit{Network} chips (\textit{HICANNs}, \citealp{schemmel_ijcnn2008}), that provide a highly configurable substrate which physically emulates adaptively spiking neurons and dynamic synapses.
The intrinsic time constants of these VLSI model circuits are multiple orders of magnitude shorter than their biological originals.
Consequently, the hardware model evolves with a speedup factor of $10^3$ up to $10^5$ compared to biological real time, the precise value depending on the configuration of the system.

In addition to a high-bandwidth asynchronous on-wafer event communication infrastructure, full custom digital off-wafer ASICs provide terminals for a packet-based multi-purpose communication network \citep{SCHO10a}.
These so called \textit{Digital} \textit{Network} \textit{Chips} \textit{(DNCs)} are backed by a flexible FPGA\footnote{Field Programmable Gate Array} design that handles the packet routing \citep{HART10}. 
The communication infrastructure is illustrated in Figure~\ref{figure:stage2_comm_structure}.
See Section~\ref{section:wafer_communication} for details on the inter-chip communication scheme.

A full wafer system will comprise $384$ interconnectable HICANNs, each of which implements more than 100,000 programmable dynamic synapses and up to 512 neurons, resulting in a total of approximately 45 million synapses and up to 200,000 neurons per wafer.
The exact number of neurons depends on the configuration of the substrate, which allows to combine multiple neuron building blocks to increase the input count per cell. 

\subsubsection{Composition of the FACETS Hardware System}
\begin{figure}[htb]
    \centering		
    \includegraphics[width=\columnwidth]{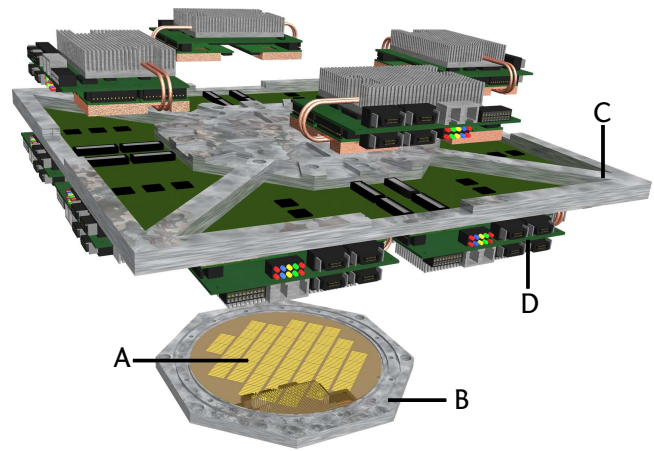}
    \caption{The FACETS wafer-scale hardware system: Wafer (A) comprising HICANN building blocks and on-wafer communication infrastructure, wafer bracket (B), top frame (C) and digital inter-wafer and wafer-host communication modules (D).}
    \label{figure:waferscale_system}
\end{figure}

The wafer as the main component for the FACETS wafer-scale hardware system has to be embedded into a framework that provides the electrical integration as well as the mechanical stability. 
The wafer has a diameter of \unit[20]{cm} and will be placed into an aluminum plate which also serves as a heat sink. 
A multi-layer Printed Circuit Board (PCB) is placed on top of the wafer. 
This PCB has to provide the fan-out of 1500 impedance-controlled differential pairs and - in the worst case - has to deliver a total electrical power of 1000 Watts to the wafer. 
A 14-layer fine pitch board with laser drilled micro-vias and a total size of \unit[430]{mm}~x~\unit[430]{mm} meets these requirements.
The PCB will be clamped to an aluminum frame that is also used as a platform for communication devices such as the 48 DNCs and the 12 FPGA boards (see Section~\ref{section:wafer_communication}). 
Figure~\ref{figure:waferscale_system} shows a 3-D drawing of the hardware composition. 
All depicted electrical and mechanical components are custom-made by FACETS project members.

\subsubsection{The HICANN Building Block}
\label{section:hicann}
\begin{figure}[htb]
    \centering		
    \includegraphics[width=\columnwidth]{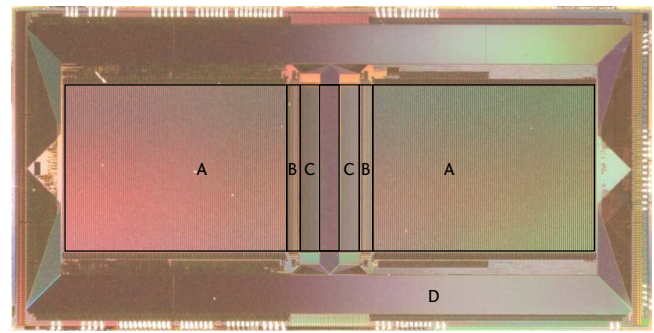}
    \caption{A photograph of the HICANN building block with synapse arrays (A), neurons (B), floating-gate arrays (C) and L1 routing (D).}
    \label{figure:hicann}
\end{figure}

The HICANN building block shown in Figure~\ref{figure:hicann} is the neuromorphic ASIC of the FACETS wafer-scale hardware system. 
The inter-chip communication scheme is explained in Section~\ref{section:wafer_communication}.

Simplifying, the HICANN can be divided into four parts: the neuron circuits with their analog parameter storage based on floating gate technology \citep{lande96}, an array of $114688$ dynamic synapses and the \emph{Layer~1 (L1)} bus system interconnecting HICANNs on a wafer.
The hardware neurons implemented by the HICANN building blocks \citep{millner10} can emulate the adaptive exponential integrate-and-fire neuron model (\emph{AdEx}, \citealp{brette_05}) which can produce complex firing patterns observed in biology (see e.g.\ \citealp{markram04, ALAN09}), like spike-frequency-adaptation, bursting, regular spiking, irregular spiking and transient spiking, by tuning a limited number of parameters \citep{naud08}.
The decision to implement this particular neuron model in hardware was motivated by the large spectrum of possible and biologically relevant cell behavior realizable with a comparably compact circuitry. 
The latter fact is a crucial aspect when aiming at the integration of large numbers of neurons in one hardware system. 
For a neuromorphic implementation of Hodgkin-Huxley cells that consume significantly more chip area see e.g.\ \cite{daouzli08weights}.

The model can be described by the following two differential equations for the membrane voltage $V$ and the adaptation variable $w$ and a reset condition specified further below:
\begin{eqnarray}
    \label{eqn:adexv}-C_{\rm{m}}\frac{dV}{dt}&=& g_{\rm{l}}(V-E_{\rm{1}})-g_{\rm{l}}\Delta_{\rm{t}}e^{\left(\frac{V-V_{\rm{t}}}{\Delta_{\rm{t}}}\right)}+w\nonumber\\
        &+&g_{\rm{e}}(t)(V-E_{\rm{e}})\nonumber\\
        &+&g_{\rm{i}}(t)(V-E_{\rm{i}}) \quad ,\\
      \label{eqn:adexw}-\tau_{\rm{w}}\frac{dw}{dt}&=&w-a(V-E_{\rm{l}}) \quad.
    \end{eqnarray}
$C_{\rm{m}}$, $g_{\rm{l}}$, $g_{\rm{e}}$ and $g_{\rm{i}}$ are the membrane capacitance, the leakage conductance and the conductances for excitatory and inhibitory synaptic inputs, where $g_{\rm{e}}$ and $g_{\rm{i}}$ depend on time and on the inputs from other neurons. 
$E_{\rm{l}}$, $E_{\rm{i}}$ and $E_{\rm{e}}$ are the leakage reversal potential and the synaptic reversal potentials. 
The parameters $V_{\rm{t}}$ and $\Delta_{\rm{t}}$ are the effective threshold potential and the threshold slope factor. 
The time constant of the adaptation variable is $\tau_{\rm{w}}$. The adaptation parameter $a$ has the dimension of a conductance.

If the membrane voltage crosses a certain threshold voltage $\Theta$, the neuron is reset: 
\begin{eqnarray}
    \label{eqn:reset}V&\rightarrow& V_{\rm{reset}} \quad ,\\
    \label{eqn:b}w&\rightarrow& w + b \quad.
\end{eqnarray}
The parameter $b$ is responsible for spike-triggered adaptation.

A neuron can be constructed out of up to 64 so-called \emph{denmem} circuits, each implementing the dynamics of the AdEx model and being connected to up to 224 synapses.
This way a neuron could have synaptic inputs from up to 14,336 other cells. 
Additionally, depressing and facilitating mechanisms of short-term synaptic dynamics (for a review see \citealp{zucker02stp}) are implemented.
A purely chip-based FACETS hardware implementation of this feature is described and applied in \cite{bill2010compensating}.

A general limitation of neuromorphic implementations of cell models is the fact that configurable parameter values will always have limited ranges.
The value ranges of all AdEx parameters configurable in the hardware implementation have been designed such that the complete set of biologically relevant firing patterns distinguished e.g.\ in \cite{naud08} can be reproduced.
If this design goal was fully achieved is currently studied with HICANN prototypes, i.e.\ work in progress (see also Section~\ref{section:calibrationresults}).

\paragraph{Hebbian Learning in the FACETS Hardware} 
\label{section:hardware_stdp}

Long-term Hebbian learning in the FACETS hardware devices is implemented in every synapse as spike-timing-dependent plasticity (STDP, reviewed e.g.\ in \citealp{morrison08_stdp}).
To ensure high flexibility in terms of mappable neuronal networks each neuron in hardware needs an appropriate number of synaptic inputs.
However, due to limited die area, a trade-off between the number of synapses and the chip resources for a single synapse has to be made.

To achieve a minimal circuit size for the synapses, local correlation measurements and the local synaptic weight storage are separated from global weight \emph{update controllers} \citep{schemmel_iscas07, schemmel_ijcnn06}.
Causal and acausal correlations between pre- and post-synaptic spikes determine the temporal factor of the STDP rule described in \citet{schemmel_ijcnn04} and are accumulated locally until they are processed by the update controller.
Synaptic weights are stored locally as digital values with a four-bit resolution each.
This resolution is again a trade-off between precision and chip resources and requires several correlated events to reach the next discrete weight value.
If a sufficient amount of correlations is accumulated, the discrete weight is updated by the update controller.
Since many synapses share one update controller a weight update is performed periodically with a frequency that has an upper limit determined by the circuitry \citep{schemmel_ijcnn06}.
Since a reduced symmetric nearest-neighbor spike pairing scheme turned out to be one feasible approach for describing biological measurements \citep{burkitt07, morrison08_stdp}, this specific plasticity mechanism has been chosen to be implemented in hardware to further reduce the size of a synapse.
Update controllers are modifying the synaptic weights by using look-up tables that are listing, for each discrete weight value, the resulting weight values in case of measured causal or acausal correlations.
These look-up tables can be adapted to the weight-dependent factor of any STDP rule.

Despite its global weight update controllers, the STDP mechanism of the FACETS hardware has to be considered local to every synapse.
The implementation of this particular model represents a project-wide decision on the most promising mechanism to be cast into silicon, taken in the early phase of FACETS.
Recent developments in the modeling of learning and self-organization in neural networks (see e.g.\ \citealp{sjostrom2008dendritic, pfeiffer2010reward}) combine such local rules with various global mechanisms like the reward-based modulation of large groups of synapses.
With respect to more complex and relevant plasticity mechanisms, an extension to STDP rules with additional input parameters, e.g.\ membrane potentials, spike rates or global reward signals, is currently under development.

\paragraph{Parameter Memories}
\label{section:param_memory}

In contrast to most other systems, the FACETS wafer-scale hardware deploys analog floating gate memories similar to cells developed by \citet{lande96} as storage devices for the analog parameters. 
Due to the small size of these cells, most parameters can be provided individually for a single neuron circuit. 
This way, matching issues can be counterbalanced, and different types of neurons can be implemented on a single chip.

As a starting point for the parameter ranges, parameters from \citet{brette_05} and \citet{destexhe1998} have been used. 
The chosen ranges allow leakage time constants $ \tau_{\rm{mem}}=C_{\rm{m}}/g_{\rm{l}}$ at an acceleration factor of $10^4$ between \unit[1]{ms} and \unit[588]{ms} and an adaptation time constant $\tau_{\rm{w}}$ between \unit[10]{ms} and \unit[5]{s} in terms of biological real time. 
The parameters used by \cite{pospischil08}, for example, lie easily within this range.

\begin{figure}[htb]
    \centering		
    \includegraphics[width=.75\columnwidth]{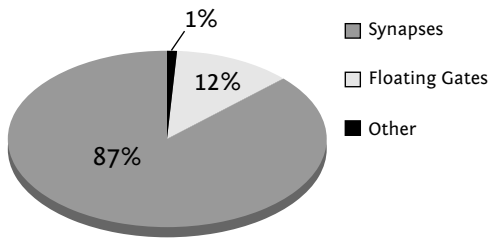}
    \caption{Sector diagram of the parameter space to configure one HICANN chip. For a full wafer, the configuration data volume is \unit[44]{MB} large.}
    \label{figure:memsector}
\end{figure}

A substantial amount of digital memories is integrated in the chip, dominated by the synapse RAM. 
Each of the $114,688$ synapses has 8 bit memory cells for weight and address storage. 
For the whole wafer, the synapse RAM alone is \unit[38]{MB} large. 
Figure \ref{figure:memsector} shows the partitioning of the parameter memory on a HICANN building block. 
To compare the analog floating gates to normal digital memory, each cell has been counted as \unit[10]{bit}, since this is the number of bits needed to program it. 

\begin{figure*}[htbp]
    \centering
    \includegraphics[width=.95\linewidth]{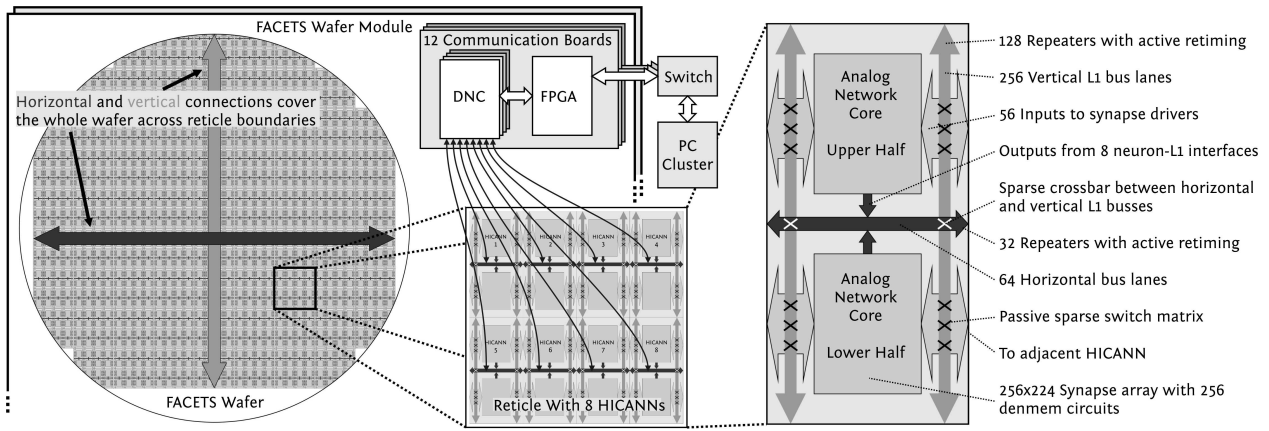}
    \caption{\label{figure:stage2_comm_structure}Communication structure on a wafer module of the FACETS wafer-scale hardware system. Neural activity is transported horizontally (dark gray) and vertically (light gray) via asynchronous L1 buses on the HICANN building blocks. Repeater circuits at the edges of these blocks allow for a distribution of the buses over the whole wafer. Off-wafer connectivity is established by the L2 network via DNCs and FPGAs. It interfaces the L1 buses on the HICANN building blocks. Several wafer modules can be interconnected using routing functionality between the FPGAs via Ethernet switches.}
\end{figure*}

\subsubsection{Communication Infrastructure}
\label{section:wafer_communication}

The communication infrastructure of the FACETS wafer-scale hardware is illustrated in Figure~\ref{figure:stage2_comm_structure}. 
Pulse communication is generally based on the digital transmission of neural events representing action potentials, but a distinction in two network layers can be made.
An asynchronous, serial protocol, named \textit{Layer 1 (L1)} utilized by HICANNs at a wafer level provides intra-wafer action potential transmission on a high density direct interconnection grid.
A second one, named \textit{Layer 2 (L2)}, deploys the DNCs and FPGAs for synchronous, packet-based, intra and inter-wafer communication and - compared to L1 - establishes a more flexible routed network of lower density. 
To cope with inevitable jitter in routing delay, a time stamp is transmitted together with the address within the data packets of this network.
A PC cluster that handles the mapping, configuration and control process described in Section~\ref{section:software} as well as the playback and recording of external stimuli to the neural network is connected to the FPGAs via multi-Gigabit Ethernet.

Activity is injected into the L1 network in the form of 6\,bit serial pulse address packets by neurons that connect to the horizontal buses. 
Sparsely populated passive switch matrices at the intersections of horizontal and vertical buses pass the data to the vertical buses. 
Further sparse switch matrices connect to horizontal lines feeding synapse drivers that act as data sinks to the network. 
While crossing HICANN block boundaries the signals are refreshed by repeater circuits with active re-timing that are capable of driving the signals across one HICANN block. 
The sparseness of the switch matrices is chosen such that the repeater circuits are not overloaded while still providing maximum flexibility for implementing various neural network topologies (see \citealp{fieres_ijcnn2008} and \citealp{schemmel_iscas2010} for more information on the underlying design decisions and analyses of the resulting limitations).

Connectivity between the HICANN blocks is established by edge connecting them in the layout. 
As illustrated in Figure~\ref{figure:stage2_comm_structure}, this is only possible for eight HICANNs located within one reticle. 
A reticle is the largest producible unit on the wafer and no connections can be formed between reticles during standard CMOS fabrication. 
Wafer-scale connectivity is obtained using a post-processing method developed in the FACETS project. 
It offers two additional routing layers that can cover the whole wafer. 
By means of this technique, an inter-reticle connection pitch well below \unit[10]{$\upmu$m} can be achieved which facilitates the required connectivity. 
Furthermore, large landing pads are formed by the post-processing that connect the wafer to the system PCB via elastomeric stripe connectors (see Figure~\ref{figure:waferscale_system} and \citealp{schemmel_iscas2010}).

These stripe connectors are used to deliver all required power to the wafer. 
Additionally, they connect high speed communication signals between the HICANNs and the DNCs\footnote{For completeness it should be noted that also analog signals, e.g.\ selectable neuron membrane voltages, are transported through the stripe connectors.}. 
This high speed communication interface transports configuration data as well as the above-mentioned L2 data packets. 
L2/L1 protocol conversion is performed inside the HICANN blocks, where L2 activity can either be injected to or read from the L1 network (see Figure~\ref{figure:stage2_comm_structure}). 
The transport of the L2 packets is handled by the DNCs, which also implement a time-stamp based buffering and heap-sort algorithm~\citep{SCHO10a}. 
Together with routing logic inside the FPGAs, the DNC-FPGA L2 network fulfills the QoS\footnote{Quality of Service} demands~\citep{philipp09QoS} for spiking neural networks, i.e.\ a constant delay at a low pulse loss rate. 
This is also true for inter-wafer connections routed through Ethernet switches connected to the FPGAs.

\subsubsection{Host Interface}
\label{section:hostinterface}

The packet communication between 
wafer and host computer passes through several layers:
DNCs, FPGA controller boards and a Gigabit Ethernet layer \citep{norris02} have to be
traversed. As each of the twelve FPGA controller boards (see C in Figure~\ref{figure:waferscale_system}) comprises two Gigabit ports
dedicated for host communication, a total bandwidth of \unitfrac[24]{GBit}{s} can
be achieved. 
Standard networking switches concentrate these links into the required number of 
\mbox{10GBase-LX4} \citep{horak03} %
upstream ports. 
A standard PC cluster equipped with
adequate network interface cards handles the traffic.
A custom design ARQ\footnote{Automatic Repeat reQuest}-style \citep{RFC3366} protocol
provides a reliable communication channel between the host computer and the
hardware system.
The FPGA controller boards act as remote terminals for these ARQ communication channels, but also provide system control functionality.

During experiments, most
communication data -- basically spike events -- flow directly between host computer and FPGA
controller boards. 
In contrast to this, in the initial setup stage almost all traffic
-- i.e.\ the system configuration data -- is dedicated to wafer communication. In this case, the
FPGA controllers act as simple transmission nodes between host computer and
wafer. Both operational stages impose
high demands on the communication bandwidth.
The initial configuration space consumes around $\unit[50]{MB}$ (see Figure~\ref{figure:memsector}).
Every spike event is represented by a 15-bit time stamp and a 12-bit data field, comprising both DNC and HICANN identifiers.
Thus, during an experiment approximately $\unitfrac[1]{GEvent}{s}$ can be transported to and from the host computer.
At a speedup factor of $10^4$, the corresponding total spike rate in the biological time domain is \unit[100]{kHz} per wafer.

To meet these requirements set by the hardware scale, acceleration factor and
modeling constraints, a highly scalable software implementation of the
communication protocol was developed
(see Section \ref{subsubsec:hw-lowlevel-interfacing} and \citealp{schilling10diplomathesis}).
This multi-threaded protocol stack already provides a zero-copy API\footnote{Application Programming Interface} to the
upper software layers.

Furthermore, to support future applications, such as interfacing the FACETS hardware system to
simulated environments which provide sensor output related to motor input, low
round-trip times between these components are crucial.
Such classes of in-the-loop experiments demand low latency communication and
high bandwidth at the same time.

\subsubsection{Chip-Based Neuromorphic System}
\label{section:stage1hardware}

On the development path towards the FACETS wafer-scale hardware platform, a purely chip-based neuromorphic system has been designed and built \citep{schemmel_ijcnn06, schemmel_iscas07} and is in active use \citep{kaplan_ijcnn2009, bruederle09establishing_hack, bruederle10simulator, bill2010compensating}.
It implements time-continuous leaky integrate-and-fire cells with conductance-based synapses and both a short-term and a long-term plasticity mechanism as described above for the wafer-scale device.
Up to 16 of these ASICs, each of which provides 384 neurons and $10^5$ configurable and plastic synaptic connections, can be operated individually or interconnected via a supporting backplane board.
This board is connected via a single Gigabit Ethernet link to a host computer, through which multiple users can access and use the neuromorphic devices in parallel.
The possibility of remotely accessing the chips via the Internet in addition to setting up and running experiments with an available PyNN interface (see Section\ \ref{section:pynn} and \citealp{bruederle09establishing_hack}) already now make this system a tool that is used for neuromorphic model exploration by users from various countries.
Many circuit design strategies for the wafer-scale system are implemented for testing purposes in this chip-based device, including the STDP correlation measurements (see Section~\ref{section:hardware_stdp}) located in every individual synapse. 
Basic plasticity studies supporting the design of the wafer-scale system, some of which are outlined in Section~\ref{stdp_methods}, incorporate investigations on the basis of experimental results from the chip-based devices.

\subsection{Software Framework}
\label{section:software}

\begin{figure}[htb]
    \centering
    \includegraphics[width=.95\columnwidth]{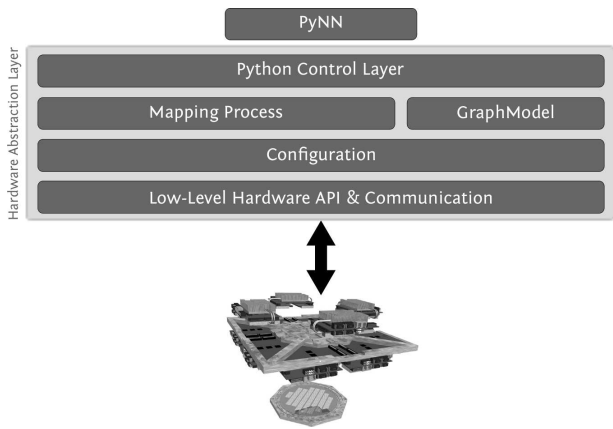}
    \caption{\label{figure:software_stack_goal} Schematic of the \emph{hardware abstraction layer}, i.e.\ the stack of software modules for the automated and bidirectional translation between PyNN model descriptions and appropriate hardware configuration and control patterns. The individual modules are: A Python control layer, a mapping layer that operates on a graph-based data container (\emph{GraphModel}), and low-level layers that deliver the generated hardware configuration patterns and control sequences via a dedicated communication protocol.}
\end{figure}

Figure~\ref{figure:software_stack_goal} shows the stack of software modules that will be described in the following.

Its components seamlessly interact in performing an automated translation of arbitrary neural network experiment descriptions into appropriate data for hardware configuration and control.
The same stack also automatically re-interprets the acquired hardware output into its biological interpretation.
The top-level interface offered to hardware users to describe neural network setups is based on the simulator-agnostic modeling language PyNN. 
The concept of this approach and its advantages, especially for neuromorphic system operation, will be described in Section~\ref{section:pynn}. 

The process of mapping a PyNN description onto the configuration space of the FACETS hardware systems, including dedicated representation formats, will be described in Sections~\ref{sec:mapping_process} to \ref{sec:multichip}.
Sections~\ref{section:mapping_analysis} and \ref{section:virtual_hardware} focus on the mapping analysis plus its testing and optimization on the basis of an elaborate virtual version of the wafer-scale hardware system.
The special performance requirements for the low-level host-to-hardware communication software and the implemented corresponding solutions are outlined in Section~\ref{subsubsec:hw-lowlevel-interfacing}.

\subsubsection{PyNN \& NeuroTools}
\label{section:pynn}

\begin{figure*}[htb]
    \centering
    \includegraphics[width=.95\linewidth]{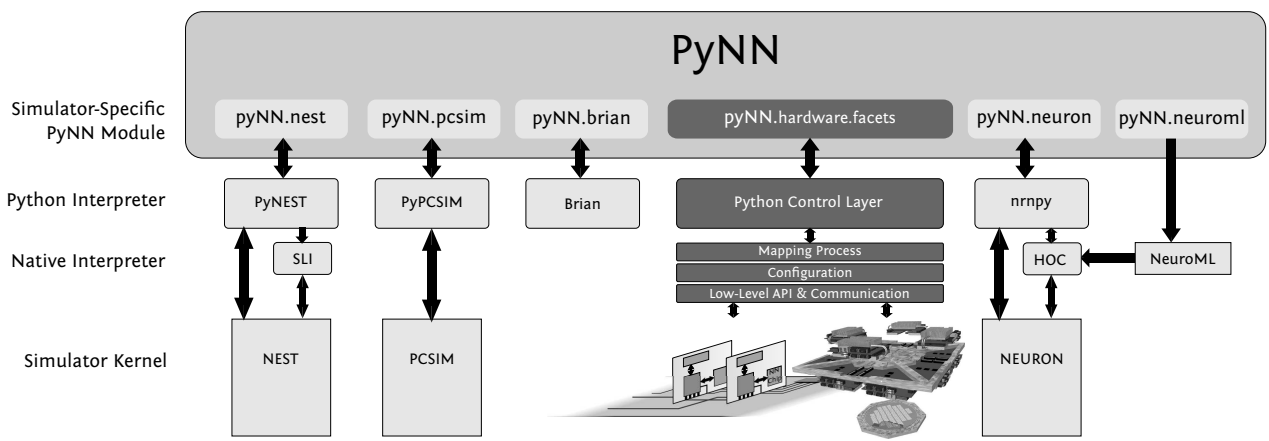}
    \caption{\label{figure:pynn_schematic} Schematic of the simulator-independent modeling language PyNN. Like various established software simulators, the FACETS neuromorphic hardware systems have been integrated into the PyNN unification and standardization concept.}
\end{figure*}

PyNN is a simulator-independent, Python-based language designed for describing spiking neural network models \citep{davison08pynn}.
It offers functions and classes for the setup and control of experiments, and it provides standard cell models as well as standardized dimension units.
PyNN supports various software simulators like NEURON \citep{hines06neuron,hines09}, NEST \citep{scholarpedia_articlenest,eppler2008}, Brian \citep{goodman2008} and PCSIM \citep{pecevski09}.
With PyNN, which is open source and well documented, a user can set up a neural network model, run it on any of the supported back-ends without changing the code, and directly compare the results.
This provides the possibility to conveniently port experiments between different simulators, to transparently share models and results, and to verify data acquired from different back-ends (see Figure~\ref{figure:pynn_schematic}).

The integration of the operating software framework for the FACETS hardware system into the PyNN concept \citep{bruederle09establishing_hack, bruederle09phd} is a crucial aspect of the presented neuromorphic workflow.
One important motivation for this approach is to create a bridge between the communities of neuromorphic engineers and neural modelers, who have been working in rather separate projects so far.
The back-end agnostic concept of PyNN, now also offering the possibility to port existing experiments between the supported software simulators and the FACETS hardware system, allows to benchmark and verify the hardware model. 
The API of PyNN is easy to learn, especially for scientists who have already worked with software simulators.
Hence, PyNN represents an optimal way to provide non-hardware experts a convenient interface to work with the FACETS neuromorphic devices.
In general, PyNN interfaces to neuromorphic systems make it possible to formulate transparent tests, benchmarks and feature requests, and therefore can influence and boost biologically oriented hardware development.
They might, eventually, support the establishment of such emulation devices as useful modeling tools. 

On top of PyNN, a library of analysis tools called NeuroTools \citep{neurotools_homepage} is being developed, which builds upon the interface and data format standards, but also exploits the possibility to incorporate third-party Python modules e.g.\ for scientific computing and plotting \citep{Oliphant2007,Scipy2001,langtangen_python, Matplotlib2007}.
Thus, for all supported software simulators and for the FACETS neuromorphic hardware systems, all stages of neural modeling experiments - description, execution, result storage, analysis and plotting - can be performed from within the PyNN and NeuroTools framework.

\paragraph{Simulations as Reference for Translation and Calibration}
The hardware-specific PyNN approach incorporates quantitative bidirectional translation methods between the neuromorphic system dynamics and the biological domain, both in terms of electrical variables and the different time domains.
This translation incorporates calibration routines that minimize the impact of transistor-level fixed-pattern noise on the behavior of neural and synaptic circuits.
The translation and calibration scheme developed for the FACETS hardware systems directly involves reference software simulations for the biologically relevant gauging of hardware parameters, heavily exploiting the PyNN paradigm of unified setup descriptions. 
Section~\ref{section:param_translation} provides more details on this.

\subsubsection{Mapping Process}
\label{sec:mapping_process}

\begin{figure}[htb]
    \centering
    \includegraphics[width=\columnwidth]{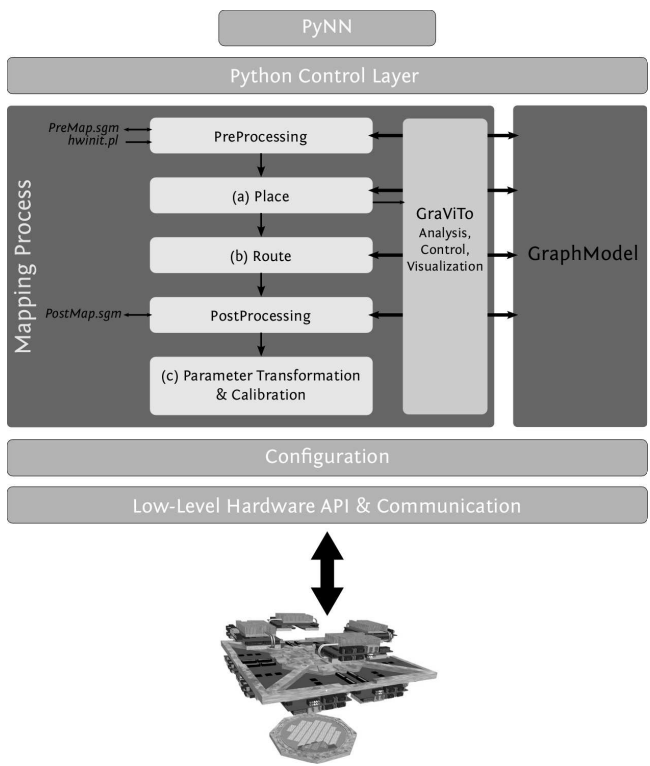}
    \caption[Mapping PyNN neural network model descriptions onto the configuration space of the wafer-scale hardware system]{Mapping PyNN neural network model descriptions onto the configuration space of the wafer-scale hardware system. The three main processing steps, all operating on one unified data container (GraphModel), are 
    \begin{inparaenum}[(\bf a\rm)]
        \item the \emph{placing} of neurons onto the available circuitry, 
        \item the realization of synaptic connections by appropriately configuring the available \emph{routing} infrastructure on the device, and
        \item the transformation of neuron and synapse parameters into corresponding parameter values offered by the highly configurable device. 
    \end{inparaenum}
    The latter step can involve calibration data to tune individual sub-circuits such that the effect of unavoidable transistor-level variations onto the mapped models is minimized.}
  \label{figure:workflow}
\end{figure}

The mapping process determines a valid routing network configuration and parameter value set as initial setup data for the FACETS hardware system. 
This takes into account topology constraints between hardware blocks such as connectivity, connection counts, priorities and distances as well as source/target counts. 
Figure~\ref{figure:workflow} depicts the single steps of the mapping process as described by \cite{ehrlich2010anniip}.

The mapping is accomplished in the three main steps of \textit{Placement}, \textit{Routing} and \textit{Parameter Transformation \& Calibration}, with an appropriate \textit{Pre-} and \textit{PostProcessing} of the configuration data.
As the first three main steps are explained in more detail in the following we will shortly summarize the functionality of the remaining parts.

Starting with a neural architecture defined via PyNN, the first mapping step of \textit{PreProcessing} reads in a description of the hardware (see Section~\ref{section:hwmodel}), described using the novel query language \textit{GMPath} (see Section~\ref{section:graphmodel}).
It sets up an internal representation for both the hardware and the biological model in the form of a directed graph called \textit{Graph Model} (see Section~\ref{section:graphmodel}). 
Optionally, a so-called \textit{PreMapping} netlist of the biological model can be streamed out into a file. 
Following placement and routing, the same applies for the \textit{PostProcessing} with a \textit{PostMapping} netlist, which includes the possibility to obtain a PyNN script that represents the (possibly distorted) network ultimately realized on the hardware back-end.

The individual steps of the process are automatically initiated and partly controllable via the PyNN module for the FACETS hardware system. 
Furthermore a stand-alone software named \textit{GraViTo} is provided for the analysis of the mapping results (see Section~\ref{section:mapping_analysis}).

\subsubsection{Internal Hardware Description}
\label{section:hwmodel}
Prior to the mapping process we have to define the hardware in an abstract manner. 
For this purpose we utilize the path language \textit{GMPath} to set up an appropriate GraphModel (both described in Section~\ref{section:graphmodel}) as a versatile internal representation.

\begin{figure}[htbp]
    \centering
    \includegraphics[width=.9\columnwidth]{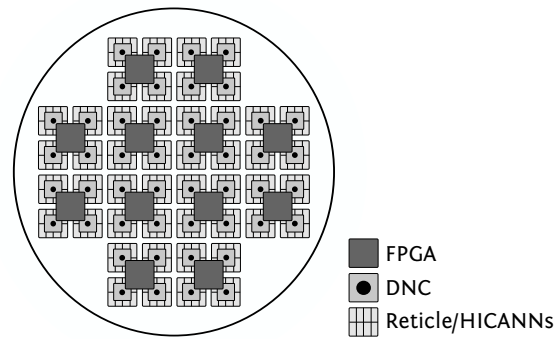}
    \caption{Example FACETS wafer-scale hardware setup from \cite{ehrlich2010anniip}: 12 FPGAs control 48 DNCs, which are connected to 384 HICANN ASICs.}
    \label{figure:architecture}
\end{figure}

In Figure~\ref{figure:architecture} a FACETS wafer-scale hardware setup - also applied in \cite{ehrlich2010anniip} - is illustrated.
As described in Section~\ref{section:hardware}, the fundamental layer of the FACETS wafer-scale hardware is an array of reticles shown as light gray squares, housing the HICANN circuitry that implements neural functionality, with a second layer of DNCs above. 
The third and topmost layer represents a regular grid of FPGAs, colored dark gray. 

\subsubsection{The GraphModel Container}
\label{section:graphmodel}

A data model called \emph{GraphModel} \citep{wendt08} represents both the targeted biological and the configurable hardware structure within the mapping software.
It can be characterized as a hierarchical hyper graph and consists of vertices (data objects) and edges (relationships between the vertices). 
A vertex contains a single data value.
An edge can be one of the following types:
\begin{description}[hierarchical:]
	\item[hierarchical:] models a parent-child relationship, structuring the model
	\item[named:] forms a directed and named relation between any two vertices in the model
	\item[hyper:] assigns a vertex to a named edge, characterizing the edge in more detail
\end{description}
The major advantage of this graph approach are the implementation convenience and efficiency as well as the flexibility to achieve the complex requirements from both the biological and the hardware model. 
Due to the structure of the graph model it can be easily (de-)serialized, providing save and restore functionality. 
Via the path-based query-language \textit{GMPath} \citep{wendt2010anniip} information can be dynamically retrieved from and stored to the models. 
The GraphModel is used to store all information during the configuration process, i.e.\ the models themselves, the mapping, routing and parameter transformation algorithms data and their results. 

\begin{figure}[htbp]
    \centering
    \includegraphics[width=.9\columnwidth]{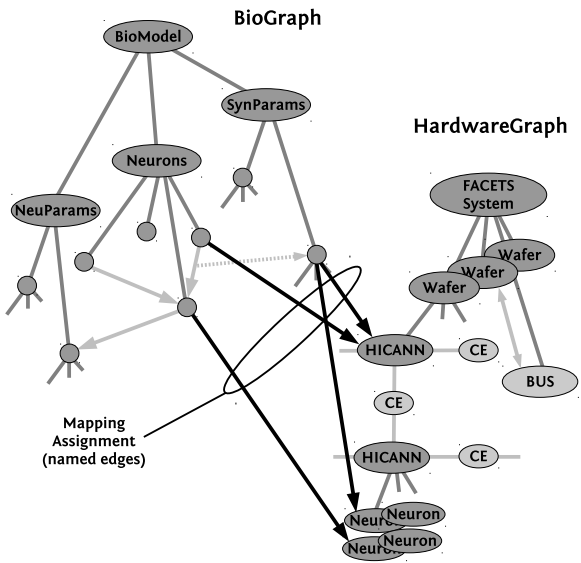}
    \caption{A simplified example of two graph models, assigning neural elements to hardware components.}
    \label{figure:MappingGraphModel}
\end{figure}

Figure \ref{figure:MappingGraphModel} shows the graph model representation of a biological network (called \emph{BioGraph}) and its hardware representation (called \emph{HardwareGraph}), connecting elements via named edges after a placement step.

\paragraph{The Query Language GMPath}
\label{section:gmpath}

To retrieve information from and propagate data to the graph models, the path-based query language \textit{GMPath} was developed, providing a universal interface for placing and routing algorithms as well as for configuration, visualization and analysis tools \citep{wendt2010anniip}.
Based on so-called navigational steps, a path request can enter the model at any point (node or edge) and addresses iteratively the logical environment by
\begin{itemize}
    \item[$\cdot$] shifting the focus hierarchically up- or downward
    \item[$\cdot$] shifting the focus back and forth along edges
    \item[$\cdot$] filtering according names
    \item[$\cdot$] concatenating sub-queries
\end{itemize}

The results are lists of nodes or edges and serve the requesting software as model information input. 
Because of its string based format and the ability to address nodes or edges unambiguously, the queries can be created conveniently and dynamically at runtime and can be used to extend and modify the models.

\begin{figure}[htbp]
	\centering
	\includegraphics[width=\columnwidth]{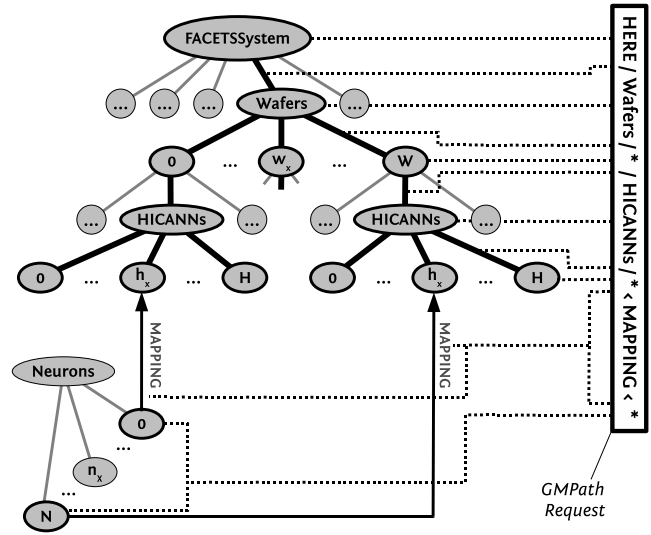}
	\caption{An example GMPath request to retrieve all assigned neurons of the biological model.}
  \label{figure:GMPathExample}
\end{figure}

Figure \ref{figure:GMPathExample} exemplarily shows subsequent navigational steps of an executed path request, which enters the abstract hardware model at its root, addresses all existing HICANN nodes and finally follows incoming mapping edges to their origins, the neurons of the biological model.

\subsubsection{Neuron Placement}
\label{section:placement}

The process of assigning neural elements like neurons, synapses or their parameters to distinct hardware elements is called \emph{placement}. 
It can be characterized as a multi-objective optimization problem, the solution of which significantly influences the overall mapping results. 
Typical algorithmic approaches create clusters of cells with common properties that are mapped to the same HICANN building blocks.
Possible optimization objectives are:
\begin{itemize}
    \item[$\cdot$] minimize neural input/output variability cluster-wise
    \item[$\cdot$] minimize neural connection count cluster-wise
    \item[$\cdot$] comply with parameter limitations
    \item[$\cdot$] comply with cluster capacities (neural capacity of hardware elements)
    \item[$\cdot$] minimize routing distances
\end{itemize}

In order to achieve these objectives with user-defined weightings in acceptable computation time, a force-based optimization heuristic was developed. 
This algorithm balances modeled \emph{forces} (special implementations of the optimization objectives) in an $n$-dimensional space until an equilibrium is reached and a final separation step assigns data objects to clusters with affine properties.
Despite this problem being NP-complete, significantly improved results can be found with this algorithm in an acceptable computation time, as compared to a fast random placement. 

\begin{figure}[htbp]
    \centering
    \includegraphics[width=.94\columnwidth]{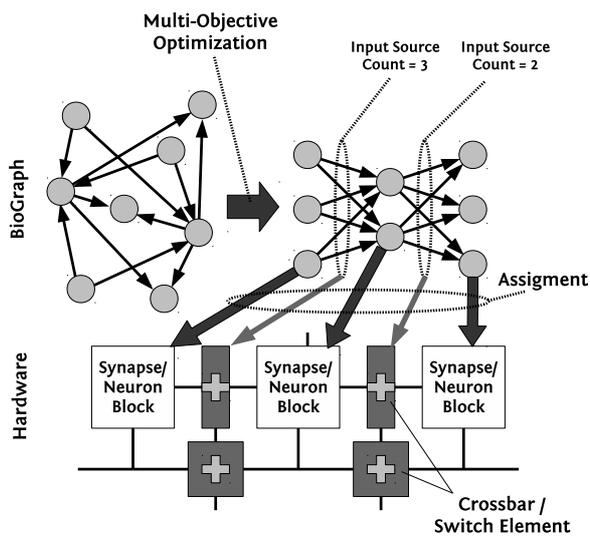}
    \caption{An example placement, divided into an optimization and assignment step. }
    \label{figure:PlacementExample}
\end{figure}

Figure~\ref{figure:PlacementExample} illustrates a placement process, divided into an optimization step, which sorts the given biological network for optimal hardware utilization with regard to the input source variability, and an assignment step, defining the physical realization of neural elements on the hardware system.

\subsubsection{Connection Routing}
\label{section:routing}

The routing step allocates and configures the hardware resources for establishing the synaptic connections in the already placed BioGraph. 
Given the fixed amount of available resources it is not evident a priori whether arbitrary network topologies are always perfectly reproducible.

Synaptic connections can in principle be established via the L1 and L2 infrastructure (see Section~\ref{section:wafer_communication}). 
In the approach described here, all intra-wafer connectivity is routed exclusively on L1. 
The L2 network is reserved for inter-wafer connections in a multi-wafer system.

The intra-wafer routing algorithms were developed in close cooperation with the wafer design~\citep{fieres_ijcnn2008}. 
Some hard-wired features of the L1 infrastructure are thus laid out to optimally suit the routing requirements. 
The routing itself is performed in two stages. 
The first stage establishes connections on a HICANN-to-HICANN level via the horizontal and vertical L1 buses, mainly by configuring the bus repeaters and sparse crossbars (see Figure~\ref{figure:stage2_comm_structure}). 
In the second stage the signals are routed from the vertical L1 bus lanes into the synapse arrays via the sparse switch matrices, the synapse drivers and the address decoders of the synapses, the latter not being shown in Figure~\ref{figure:stage2_comm_structure}.

The algorithms were proven in various test scenarios: Homogeneous randomly connected networks with up to 16,000 neurons, locally connected networks (according to \citealp{tao04egalitarian}) as well as a model of a cortical column (following \citealp{binzegger04quantitative} and \citealp{kremkow07neurocomputing}) with 10,000 neurons. 
It turns out that in typical cases only a small amount of unrouted connections must be accepted, mainly due to limited resources in the second routing stage. 
However, it was also shown that by decreasing the density of the neuron placing the routing can be generally facilitated, at the expense of a larger portion of idle hardware synapses. 

The routing algorithms proved to be also applicable for the benchmarks described later in this publication, see Section~\ref{section:benchmarks} and \cite{ehrlich2010anniip}.

\subsubsection{Parameter Transformation}
\label{section:param_translation}
The parameter transformation maps parameters of given neuron and synapse models into the hardware parameter space.
It is performed HICANN-wise.
Biological data is first acquired from the so-called \emph{BioGraph} of the GraphModel (see Section~\ref{section:graphmodel}) and then transformed into a hardware configuration, which is stored back into the \emph{HardwareGraph}.
For an adequate biology-to-hardware translation several constraints have to be considered, such as hardware imperfections and shared or correlated parameters in the microchip.

For the membrane circuits, a two-step procedure was developed to translate the 18 biological parameters from the PyNN description to the 24 electrical parameters of the HICANN building block. The first step is to scale the biological neuron model parameters in terms of time and voltage. At this stage, the desired acceleration factor is chosen and applied to the two time constants of the neuron model. Then, the biological voltage parameters are transformed to match the voltage levels of the HICANN building block. The second step is to translate those intermediate values to appropriate hardware parameters. For this purpose, each part of the membrane circuit was characterized in transistor-level simulations, which were used to establish the translation functions between the scaled AdEx parameters and their hardware counterparts.

\begin{figure}[htbp]
	\centering
	\includegraphics[width=\columnwidth]{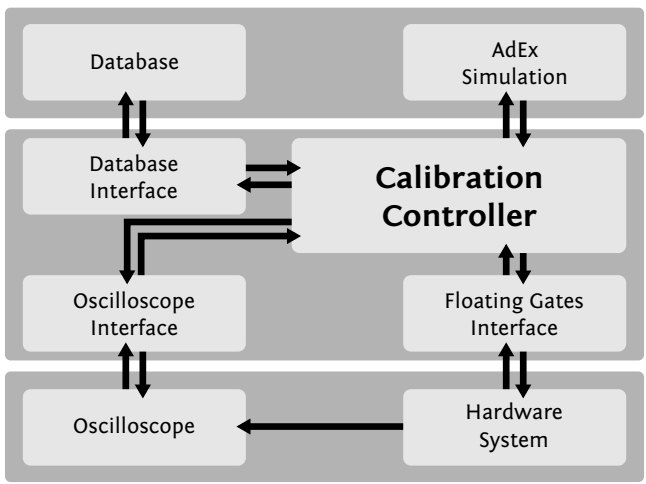}
	\caption{Architecture of the calibration software. The main component, the calibration controller, executes the calibration algorithms and communicates with the hardware, the oscilloscope and the database via dedicated interfaces. The calibration software also incorporates an AdEx model simulator to compare software and hardware results.}
        \label{figure:CalibSoft}
\end{figure}

However, due to transistor size mismatch in the hardware, these translation functions are expected to differ from neuron to neuron. A calibration software has been developed to automatically parameterize these translation functions for each neuron. For each neuron model parameter, the software will send a spectrum of values to the HICANN building block, and measure the resulting membrane potentials of the current cell. It will then deduce the corresponding AdEx parameters from these measurements, and store the value pairs into a database. After a given number of measurement points, the algorithm will compute the relation between the hardware parameters sent to the floating gates and the AdEx parameters, and store this function into the database. Figure \ref{figure:CalibSoft} illustrates the calibration software architecture.

Once the calibration step is done, the database can be used to automatically convert a set of biological neuron parameters to their hardware counterparts, allowing on-the-fly conversion of neuron parameters for the wafer-scale hardware system.

Concerning the synapses, there are mainly two restrictions ensuing from the chip design: 256 synapses of the same row share the maximal conductance $g_{\mathrm{max}}$ and the short term plasticity mechanism, and weights are restricted to a 4-bit resolution. By averaging over all active synapses, the transformation algorithm determines $g_{\mathrm{max}}$ and sets the digital weights accordingly, using \emph{stochastic rounding} to avoid systematic errors.

\subsubsection{Application of the Mapping Flow onto the FACETS Chip-Based System}
\label{sec:multichip}

In order to further demonstrate the versatility of the GraphModel-based
mapping flow introduced in Section~\ref{section:graphmodel}, we briefly outline the adoption of this
procedure to the operation of the FACETS chip-based systems (see Section~\ref{section:stage1hardware}).
This integration avoids
code redundancy by unifying the previously independent PyNN back-ends and
allows to map neural architectures onto inter-connected
chips beyond single-chip boundaries \citep{sjeltsch10diplomathesis}.
Due to the flexible design of the mapping framework, the translation of the PyNN description into the biological graph
representation (see Section\ \ref{sec:mapping_process}) and the placing of biological neurons onto their hardware
counterparts (see Section\ \ref{section:placement}) could be kept completely unchanged.
Necessary extensions were limited to the development of a new internal
hardware model that captures all features of the chip-based system as well as
adapted versions of the routing and the parameter translation (described in
Section\ \ref{section:routing} and \ref{section:param_translation},
respectively) to match the different network topology and electrical parameters. 
Together with the low-level event distribution network established by
\cite{friedmann09diplomathesis}, neural network models can now be scaled to multiple
chips.

\subsubsection{Mapping Analysis and Visualization}
\label{section:mapping_analysis}

\begin{figure}[htbp]
  \centering
	\includegraphics[width=\columnwidth]{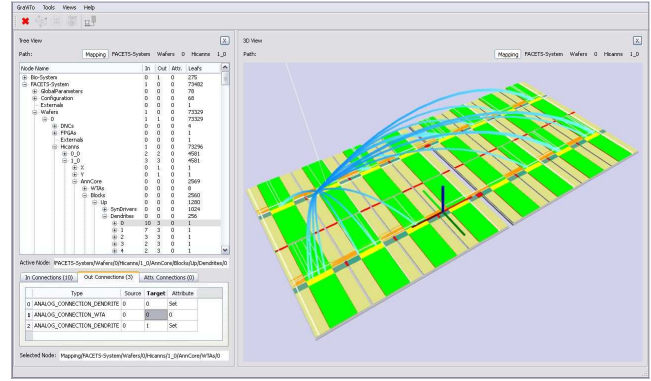}
	\caption{\label{figure:gravito}Screenshot of the GraViTo application.}
\end{figure}

The application \textit{Graph Visualization Tool -- GraViTo} as described by \cite{ehrlich2010anniip} aids in analyzing the mapping results. 
GraViTo, as shown in Figure~\ref{figure:gravito}, integrates several modules that display graph models in textual and graphical form and gathers statistical data. 
One can selectively access single nodes inside the data structure and visualize their context, dependencies and relations with other nodes in the system.

The example of the GraViTo views shows a \textit{tree view} on the left which is utilized to browse the hierarchical structure of the graph model and examine contents and connections of individual nodes.
The \textit{3-D view} on the right provides a virtual representation of the FACETS wafer-scale hardware system for interactively browsing its architecture and configuration. It also provides a global overview over the single hardware components and the networks they form.
Various statistics such as histograms for utilization of the crossbars or the synaptic connection lengths are gathered and can be displayed.

Another option for a systematic mapping analysis arises from the previously mentioned possibility to re-translate the configured HardwareGraph contents via the mapping edges through the BioGraph into a \emph{PostMapping} PyNN script.
This script intrinsically contains all model distortions caused by the mapping process, e.g.\ lost synapses and discretized or clipped parameter values.
Exploiting the PyNN concept, it can then be directly evaluated with a software simulator to extract possible functional consequences of the structural distortions, avoiding interferences with other effects like on-wafer communication bandwidth limitations.

\subsubsection{Hardware Low-Level Interfacing}
\label{subsubsec:hw-lowlevel-interfacing}
A specialized protocol of the class of \emph{selective ARQ\footnote{Automatic Repeat reQuest}
protocols} is used to
provide a fast and reliable communication channel with the neuromorphic hardware
device.
In the OSI model\footnote{Open Systems Interconnection model} this corresponds to the transport layer.

Configuration and experimental data is bidirectionally transmitted via
two \unitfrac[10]{GBit}{s} Ethernet links per FPGA.
In order to handle up to \unitfrac[2]{GByte}{s} of traffic 
while keeping the load of the host computer as low as possible, 
several software
techniques have been applied to the protocol implementation.
Various features of existing transport protocols, notably TCP, have been implemented,
including congestion avoidance, RTT\footnote{Round Trip Time} estimation and packet routing to keep the connection
in a stable and bandwidth maximizing regime.

\begin{figure}[htb]
	\centering
	\includegraphics[width=.9\columnwidth]{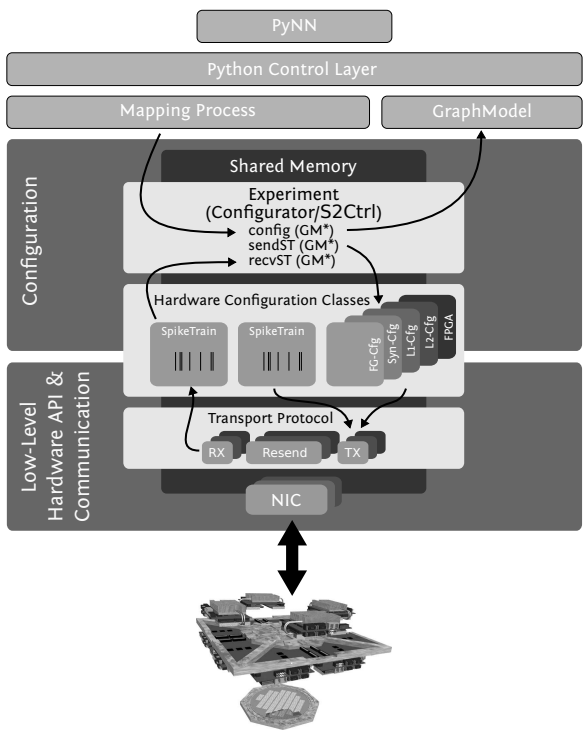}
	\caption{\label{figure:software_stack_config}Configuration and runtime control steps in the hardware abstraction layer: The \textit{Experiment} module acquires the configuration data from the Mapping Process (see Section \ref{sec:mapping_process} and Figure \ref{figure:workflow}), generates a hardware-specific representation of this data and triggers the transfer to the hardware system.}
\end{figure}

In matters of performance the framework is divided into three mainly independent processing
\emph{threads} (see Figure~\ref{figure:software_stack_config}, 
receiver thread \emph{RX}, sender thread \emph{TX} and \emph{resend} thread) 
to exploit the speed-up in execution of modern multiprocessor
systems.
Performance critical data, e.g.\ spike data can be placed in \emph{shared memory}
and passed to and from the hardware abstraction layers such as to avoid
unnecessary copying.
These shared data structures have to be protected against concurrent accesses
which imposes an additional overhead in processing time.
Thus, to keep the number of system calls and context switches small, access to data
located in shared memory is protected by means of custom built user-space fencing and locking methods.

A purely software-stack-based test has been developed that establishes a reliable ARQ connection between two host computers via 10 Gigabit Ethernet.
With a hardware-specific version of this protocol, i.e.\ with frame sizing and protocol window size, it delivers \unitfrac[10]{GBit}{s} \citep{schilling10diplomathesis}.

\subsection{Virtual Hardware}
\label{section:virtual_hardware}

An executable specification of the FACETS wafer-scale hardware system serves as a versatile tool not only during device design and optimization, but also as a test bench for all involved software layers.
It is a functional model that can be used to explore the behavior and characteristics of the real wafer-scale system in its final phase of development.

\subsubsection{Implementation}
The so-called \emph{virtual hardware} is a detailed simulation of the final hardware platform and has been implemented in \CPP\slash{}SystemC \citep{vogginger10diplomathesis}.
The virtual hardware replicates its physical counterpart in all aspects regarding functionality and configuration space. 
Every module of the real hardware has its functional counterpart in the virtual device, where especially the interface and communication structures accurately correspond to the physical system. 
It implements all analog and mixed-signal modules such as AdEx neurons and dynamic synapses (depressing and facilitating), as well as all units responsible for L1 and L2 routing. 
Compared to analog and RTL\footnote{Register Transfer Level} hardware simulations, this model is tuned towards simulation speed using behavioral models of all relevant functional components. 
However, it is possible to replace individual modules by more sophisticated models, all the way down to simulating single wires on the chip.

The current implementation of the virtual hardware differs from the real hardware system in several aspects, most of them meeting efficiency considerations.
The executable system specification is not operated from a host PC but directly from higher software layers, such that the host-to-system communication is not simulated.
Furthermore, the configuration of the HICANN building block and its components is not conducted via packets received from L2, as the software implementation of the used protocol is still under development.
Instead, every HICANN obtains its configuration via direct access to the GraphModel (see Section~\ref{section:graphmodel}).
Despite these differences the virtual hardware remains a proper replica of the FACETS wafer-scale system providing equal functionality while not suffering from hardware-specific constraints like transistor-level imperfections from the manufacturing process.

\subsubsection{Analysis And Verification Based On Virtual Hardware}
With its functionality and flexibility, the virtual hardware is an essential tool for the development of the software framework operating the FACETS wafer-scale hardware
This includes the PyNN interface and the placement, routing and parameter transformation algorithms (see Sections~\ref{section:pynn} and \ref{sec:mapping_process}), which can already be tested and verified despite the real hardware not yet being available.
The development of a hardware system, which shall be useful in a neural modeling context, can be strongly supported already during its design phase by determining constraints inherent to the system architecture, such as communication bottlenecks or the effect of shared and digitized parameters. 
Their influence can be evaluated without the interference of hardware imperfections or a missing calibration.
Such studies build the basis for improvements in the hardware design or, if possible, the development of software-based corrections.
The virtual hardware can be used from PyNN-like any other supported software simulator, thereby also offering an early modeler's perspective onto the capabilities of the future FACETS wafer-scale platform.
Any PyNN-model, in particular the benchmark models described in Section~\ref{section:benchmarks}, can be applied to this setup.
Their output can later be analyzed and compared to reference software simulations, revealing the impact of hardware constraints onto the model behavior, e.g.\ the loss of certain synaptic connections during the mapping process.

\subsection{Benchmark Model Library}
\label{section:benchmarks}

We will now present a set of experiments that serve as benchmarks for the previously described mapping process. 
The setups are implemented in PyNN and have been contributed by FACETS project partners.
They not only cover various computational aspects like memory, pattern recognition, robust information propagation in networks or dynamic switching between different functional modes, but also very different structural characteristics.

\subsubsection{Layer 2/3 Attractor Memory Model}
\label{section:layer23}

The model used here remains faithful to the model of neocortical layers 2/3 in \cite{LUND06}, and in doing so retains the modularity that is the key aspect of this architecture \citep{lundqvist2010bistable}. 
It represents a patch of cortex arranged on a hexagonal topology of $N_{\mbox{\tiny{HC}}}$ hypercolumns each separated by \unit[500]{$\upmu$m}, in agreement with data from cat cerebral cortex recordings. 
Each hypercolumn is further subdivided into $N_{\mbox{\tiny{MC}}}$ minicolumns, and various estimates suggest that there are about 100 minicolumns bundled into a hypercolumn \citep{mountcastle1997columnar, buxhoeveden2002minicolumn}. 
For the default version of the Layer 2/3 Attractor Memory benchmark model, a total number of $N_{\mbox{\tiny{HC}}} = 9$ hypercolumns and a sub-sampling of $N_{\mbox{\tiny{MC}}} = 8$ minicolumns per hypercolumn has been used. 

\begin{figure}[htb]
    \centering
    \subfloat[L2/3 network architecture]{\label{figure:kth_schematic1}\includegraphics[width=.95\columnwidth]{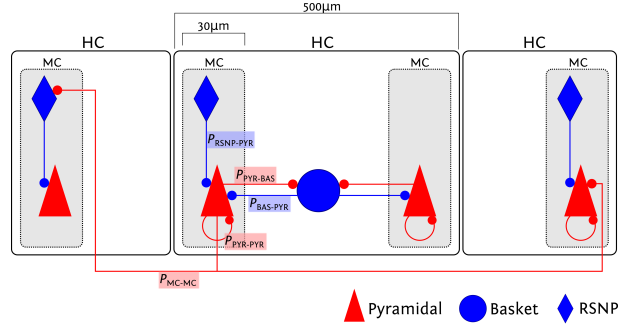}}\\
    \vspace{3mm}

    \subfloat[L2/3 model with 9 HC and 8 MC each]{\label{figure:kth_schematic2} \includegraphics[width=.6\columnwidth]{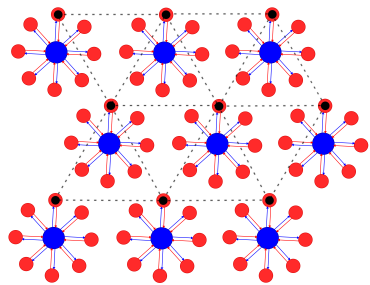}}
    \caption{\label{figure:l23_schematic} Schematic detailing the network arrangement and all the excitatory and inhibitory pathways between different cell groups and their connection densities in the L2/3 Attractor Memory network model. (a) Connectivity densities of the sub-sampled network model. See the text for further description. (b) Cartoon of a network with 9 hypercolumns (HC). Each hypercolumn has 8 circularly arranged minicolumn (MC). The large disc at the center of each hypercolumn represents a population of basket cells. Dashed lines show mutually exciting minicolumns that are distributed over different hypercolumns, forming a pattern.}
\end{figure}

The arrangement of the cells in the local microcircuit together with connection probabilities is shown in Figure~\ref{figure:kth_schematic1}. 

In the default variant of the model, each minicolumn consists of 30 pyramidal cells densely connected to other pyramidal cells in the same minicolumn ($P_{\mbox{\tiny{PYR-PYR}}} = 25\%$) and two \emph{regular spiking non pyramidal} (RSNP) cells that project to $P_{\mbox{\tiny{RSNP-PYR}}} = 70\%$ of the pyramidal cells. 
Each hypercolumn has 8 basket cells, with each pyramidal cell in a minicolumn targeting $P_{\mbox{\tiny{PYR-BAS}}} = 70\%$ of neighboring basket cells, and each basket cell targeting $P_{\mbox{\tiny{BAS-PYR}}} = 70\%$ of neighboring pyramidal cells.
The extent of basket cell inhibition is limited to its home-hypercolumn \citep{douglas04neuronal}.
Apart from these local connections, pyramidal cells located in different hypercolumns are also connected globally ($P_{\mbox{\tiny{MC-MC}}} = 17\%$).
The cartoon in Figure~\ref{figure:kth_schematic2} shows how the minicolumns in different hypercolumns, denoted by dashed lines, are connected.
We developed methods to scale this architecture up or down by means of both $N_{\mbox{\tiny{HC}}}$ and $N_{\mbox{\tiny{MC}}}$ without losing its functionality.
They are described in Section~\ref{subsubsection:Network_Scaling} and experimentally applied in Section~\ref{section:miniKTH}.

Thus, a set of mutually exciting minicolumns distributed over different hypercolumns represents a stored pattern or an attractor of the network dynamics. 
RSNP cells in a minicolumn also receive long-range excitation. 
They are excited by distant pyramidal cells, given their home minicolumn is not part of the active pattern, thus inhibiting the pyramidal cells in the minicolumn. 
In this network, we can store as many patterns as the number of minicolumns in a hypercolumn, but by allowing for overlapping memory patterns the number of stored patterns can be increased significantly. 

\begin{figure}[htb]
    \centering
    \includegraphics[width=\columnwidth]{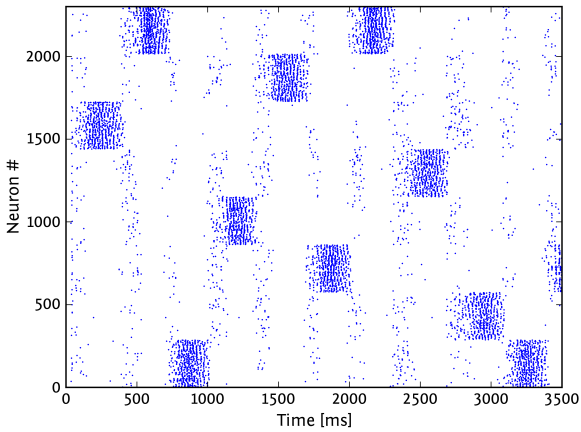}
    \caption{\label{figure:l23_rasterplot}Raster plot of characteristic activity of an L2/3 Attractor Memory network with 9 hypercolumns and 8 attractors.}
\end{figure}

Figure~\ref{figure:l23_rasterplot} shows a raster plot of the activity of the network, when all pyramidal cells are excited by individual Poisson inputs of the same rate. 

\begin{figure}[htbp]
\centering
    \subfloat[Phase space trajectory projection]{\label{figure:kth_starplot1}\includegraphics[width=.75\columnwidth]{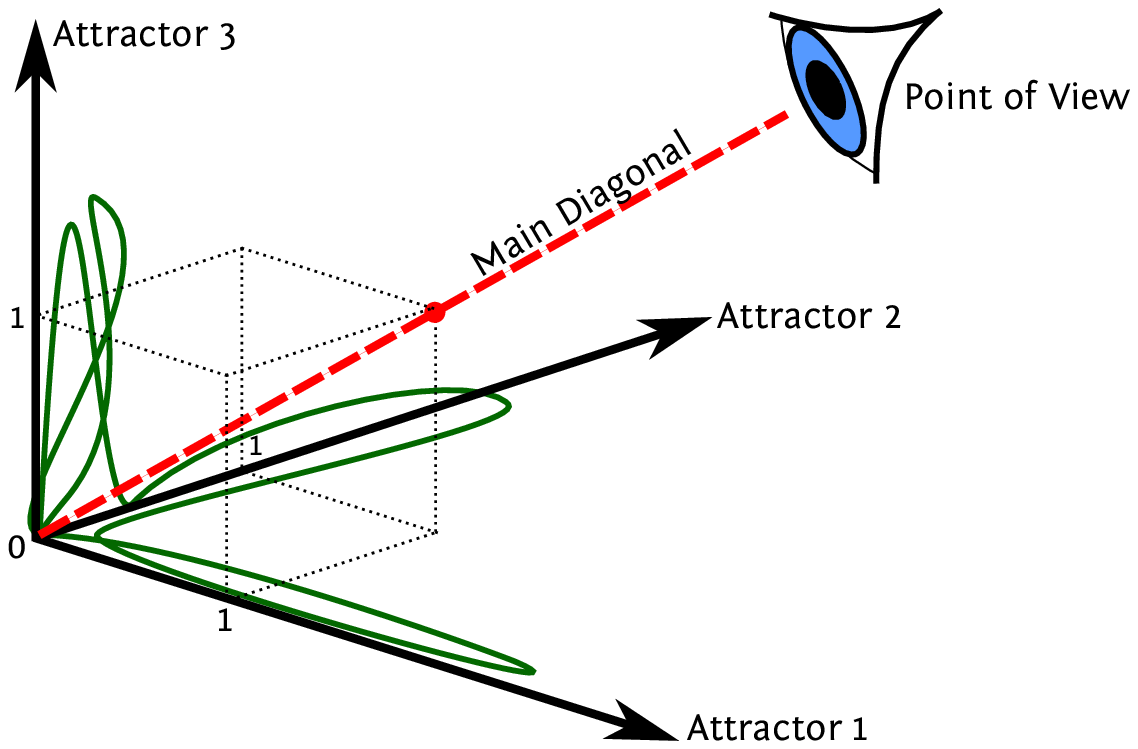}}\\
    \subfloat[Mean voltage trajectory]{\includegraphics[width=.97\columnwidth]{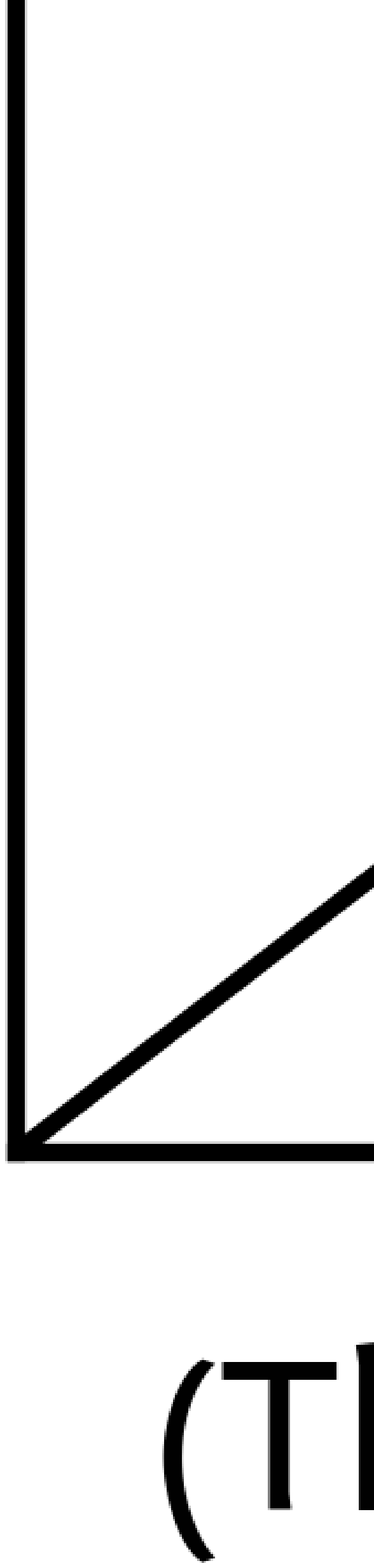}}\\
    \subfloat[Mean rate trajectory]{\label{figure:kth_starplot2}\includegraphics[width=.97\columnwidth]{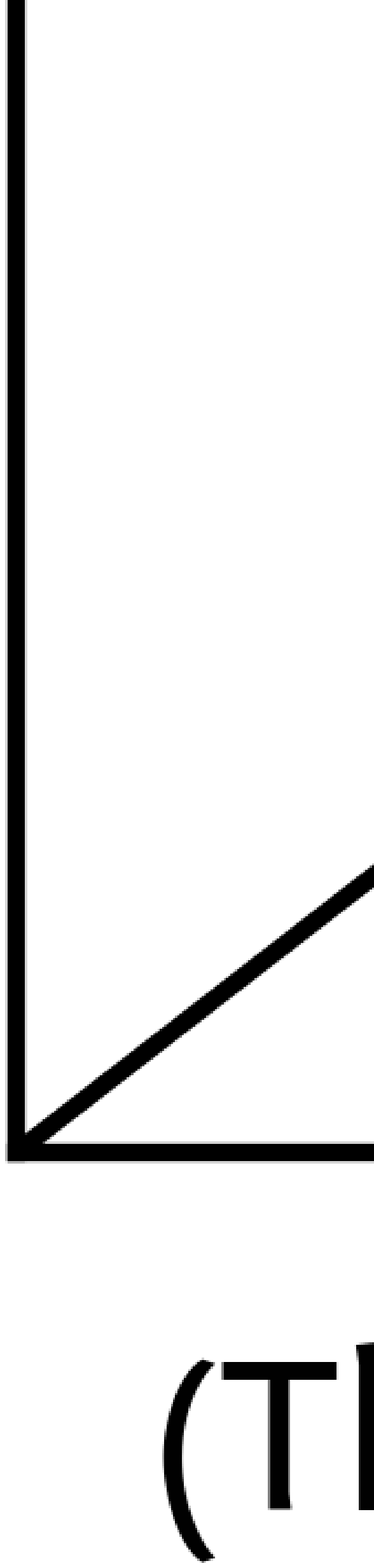}}
    \caption{\label{figure:l23_starplot}(a) Construction of phase space projection plots as shown e.g.\ in (b) and (c): The trajectory in an $n$-dimensional phase space (here: $n=3$) is projected to a hyper-plane perpendicular to the main diagonal. (b) Trajectory projection of the attractor network state evolving in 8-dimensional mean voltage and (c) mean rate phase space. Axis values represent the projected offset from a base value, which is the neuron resting potential (in \unit{mV}) for the voltage traces and $0 \unit{Hz}$ for the rate traces. The curve becomes thicker and darker as the phase space velocity decreases.}
\end{figure}
Whenever an attractor becomes stronger than the others (which happens randomly), 
it completely suppresses their activity for a short period of time. 

Pyramidal cells in an active attractor are in a so-called UP-state, where their average membrane potential is a few \unit{mV} above its rest value. 
When plotting the trajectory of the system in potential space, with each axis representing the average membrane potential of all neurons inside an attractor, a projection along the main diagonal (the line which is equidistant to all axes) will yield a typical star-like pattern (see Figure~\ref{figure:l23_starplot}). 

The synaptic plasticity mechanisms are chosen such as to prevent a single attractor from becoming persistently active. Excitatory-to-excitatory synapses are modeled as depressing, which weakens the mutual activation of active pyramidal cells in time. 
Additionally, the neurons feature an adaptation mechanism, which suppresses prolonged firing.
Both mechanisms have the effect of weakening attractors over time, such that, in contrast to a classic WTA network, also weaker patterns may become active at times.

\subsubsection{Synfire Chains}
\label{section:synfire_model}

\begin{figure}[htb]
    \centering
    \includegraphics[width=.9\columnwidth]{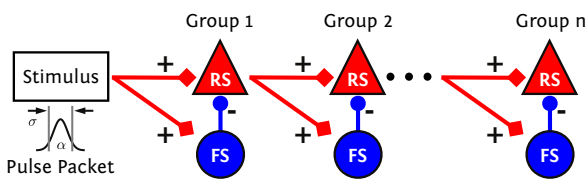}
    \caption{\label{figure:synfire_chain_schematic}Schematic of the Synfire Chain benchmark model.}
\end{figure}

Similar to classical \emph{Synfire Chain models} \citep{diesmann99, aviel03embedding, kumar08conditions, kumar10spiking}, the version chosen as a mapping benchmark consists of a chain of neuron groups connected in a feedforward fashion, with a certain delay in between. 
This allows spiking activity to propagate along the chain in a given direction (see Figure~\ref{figure:synfire_chain_schematic}). 
In addition to this, the benchmark Synfire Chain model implements feedforward inhibition by subdividing each group into a \emph{regular spiking} (RS), excitatory (80\%) and a \emph{fast spiking} (FS), inhibitory (20\%) population \citep{KREM09,kremkow10gating}. 
Inhibitory cells are also activated by feedforward projections of excitatory cells from the previous group, but project only locally onto the excitatory population of the same group with a small delay. 
This allows a fine control over the duration of spiking in a single group and prevents temporal broadening of the signal as it gets passed down along the chain. 
In the original model of \cite{KREM09}, a Synfire Chain group consists of 100 RS and 25 FS cells. 
Every cell, RS or FS, receives a total of 60 excitatory inputs from the previous RS population.
Additionally, every RS cell receives input from all 25 inhibitory neurons of the FS population within its own group.
The inhibition is tuned such that every excitatory neuron gets to spike exactly once upon activation (see Figure~\ref{figure:synfire_chain_schematic}). 

\begin{figure}[htb]
    \centering
    \includegraphics[width=\columnwidth]{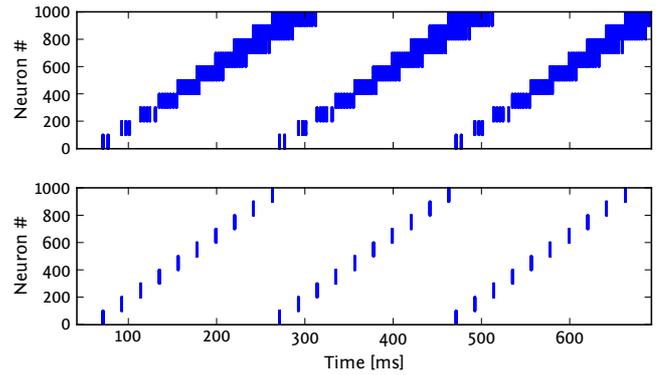}
    \caption{\label{figure:synfire_chain_activity}Raster plot of characteristic RS activity of the Synfire Chain without (top) and with (bottom) feedforward inhibition. Note the constant spike packet width in case of the active feedforward inhibition mechanism.}
\end{figure}

Methods to scale the size of this model up or down are available and described in Section~\ref{subsubsection:Network_Scaling}.
Different architecture sizes are used to benchmark the quality of the previously described mapping process. 
See Section~\ref{section:software_performance} for evaluation data based on scaled benchmark models.

\subsubsection{Self-Sustained AI States}

Randomly connected networks of integrate-and-fire neurons are known to display asynchronous irregular (AI) activity states, where neurons discharge with a high level of irregularity, similar to stochastic processes, and with a low level of synchrony \citep{brunel_jcns2000}. 
These states were also found in various other network models, including those using conductance-based \citep{vogels05signal} and nonlinear integrate-and-fire neuron models \citep{ALAN09}. 
They were shown to have properties very similar to the discharge patterns observed in awake animals \citep{elboustani07activated}.
Because cortical neurons are characterized by nonlinear intrinsic properties \citep{connors90intrinsic}, our choice of an AI state benchmark is based on the AdEx neuron model.
These nonlinear IF cells are implemented in the FACETS wafer-scale hardware (see Section~\ref{section:hicann}) and reproduce several cell classes observed experimentally in cortex and thalamus (see \citealp{ALAN09}). 

The particularity of the AI benchmark model is that it allows testing the influence of the various cell classes on the genesis of AI states by varying the different cellular properties. 
The model considers the most prominent cell classes in cerebral cortex, such as the \emph{regular spiking} (RS) cell, the \emph{fast spiking} (FS) cell, the \emph{low-threshold spike} (LTS) cell and the \emph{bursting} cells of the thalamus. 
It was found that randomly connected networks of RS and FS cells with conductance-based synaptic interactions can sustain AI states, but only if the adaptation currents (typical of RS cells) are not too strong. 
With strong adaptation, the network cannot sustain AI states. 

\begin{figure}[htb]
    \centering
    \includegraphics[width=\columnwidth]{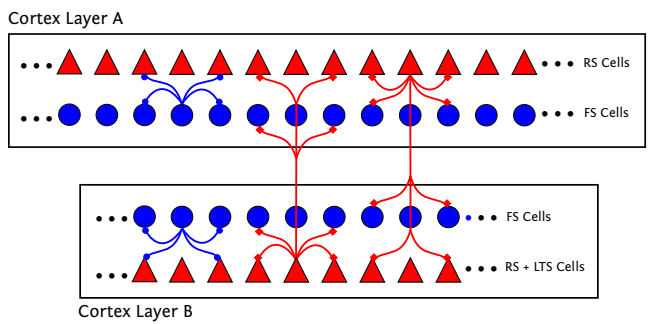}
    \caption{\label{figure:ai_schematic} Schematic of the Self-Sustained AI States benchmark model. It consists of two cortical layers A and B. Every layer has an excitatory and an inhibitory population, each of which contains certain sets of cell types (RS, FS, LTS) that determine the network dynamics (see text for details). The excitatory populations project onto every other population, while the inhibitory populations only act within their layer.}
\end{figure}

To the contrary, adding another cell class characterized by rebound responses (the LTS cell) greatly enhanced the robustness of AI states, and networks as small as about 100 neurons can self-sustain AI states with a proportion of 5\% of LTS cells. 
Interestingly, if two of such networks (one with strong adaptation, another one with LTS cells) are reciprocally connected, the resulting 2-layer network can generate alternating periods of activity and silences.  
This pattern is very similar to the Up- and Down-states observed in cortical activity during slow-wave sleep \citep{steriade2003neuronal}. 
Reducing the adaptation leads to sustained AI states, and is reminiscent of the transition from sleep to wakefulness, a sort of ``wakening'' of the network. 
In the context of this paper, we use two variants of such networks as benchmarks:
First, a network of RS, FS and LTS cells as a minimal model of AI states.  
Second, a 2-layer cortical network displaying Up and Down states.
The latter is depicted in Figure~\ref{figure:ai_schematic}.

Also this model can be scaled up and down in its size in order to benchmark the PyNN-to-hardware mapping process.
In its default version, layer A consists of 1600 excitatory RS and 400 inhibitory FS cells.
Layer B contains 400 excitatory neurons, 90\% of which are RS and 10\% of which are LTS type, as well as 100 inhibitory FS cells.
Within a single layer the connection probability is $2\%$ for a network size of 2000 cells. For smaller networks as for layer B the connection probability is rescaled inversely to the network size.
The inter-layer connectivity is excitatory only and has a connection probability of 1\%.

\subsection{Analysis Based on Software Simulations}

Compared to pure software simulators, dedicated neuromorphic hardware suffers more from limitations and imperfections, which may either directly distort the morphology of the emulated network or influence its dynamics in more subtle ways. 
On one hand, physical limitations such as size and number of implemented circuits or communication bandwidth impose rather inflexible constraints on parameters such as number of neurons and synapses or the amount of accessible data. 
On the other hand, as VLSI hardware is, inevitably, subject to manufacturing process variations, individual circuits have varying characteristics, which can only be compensated by calibration to a certain degree. 
As all these effects influence the dynamics of an emulated network simultaneously, it is usually very difficult to identify the connection between an individual cause and its effect. 
The most straightforward solution is to artificially impose individual hardware-specific distortions on software simulations, identify their impact on the network's dynamics and find, if possible, suitable compensation mechanisms.

\subsubsection{Network Scaling}
\label{subsubsection:Network_Scaling}

It is very often the case that the robustness of a network scales together with its size, or, in specific cases, with the size or number of individual components. 
However, before analyzing the effects of distortions, it is indispensable to devise a way of scaling the (undistorted) network without influencing its dynamics. 
We have developed specific rules for scaling two of our three benchmark models, in order to both explore and learn how to circumvent the limitations of the hardware.

\paragraph{Layer 2/3 Attractor Memory}
The most obvious and natural scaling of an attractor memory network lies in changing the number of attractors, i.e.\ in this particular case the number of minicolumns per hypercolumn. Also, the size of the attractors can be evenly scaled by changing the number of units per attractor, i.e.\ the number of hypercolumns. 
Finally, the size of the minicolumns itself can be scaled, by varying the number of neurons per column (excitatory and inhibitory populations can be individually tuned by changing the number of pyramidal and RSNP/basket cells, respectively).

These changes would heavily affect the network dynamics, were they not accompanied by corresponding modifications in the network connectivity. The behavior of the network is likely to remain unchanged if the excitation/inhibition patterns of each neuron are kept intact. This is most easily accomplished by keeping both the excitatory and the inhibitory fan-in of each neuron constant without modifying synaptic weights. To this end, simple scaling formulae (linear with respect to size and number of the afferent populations) for the connection probabilities between populations have been derived.

\begin{figure}[htbp]
	\centering
	\subfloat[9 HC with 3 attractors]{\label{figure:kth9x3scaling}\includegraphics[width=\columnwidth]{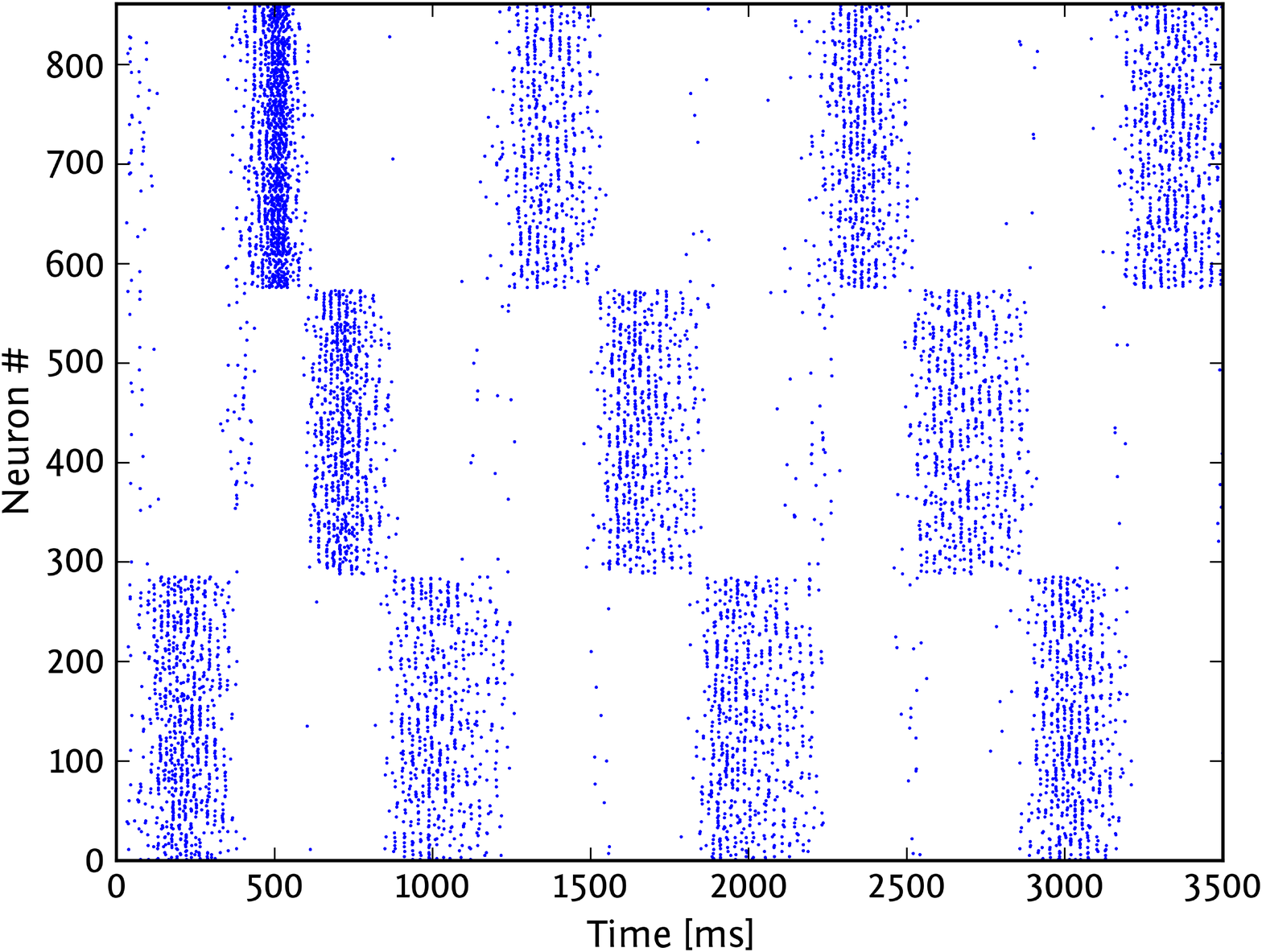}}\\
	\subfloat[8 HC with 20 attractors]{\label{figure:kth8x20scaling}\includegraphics[width=\columnwidth]{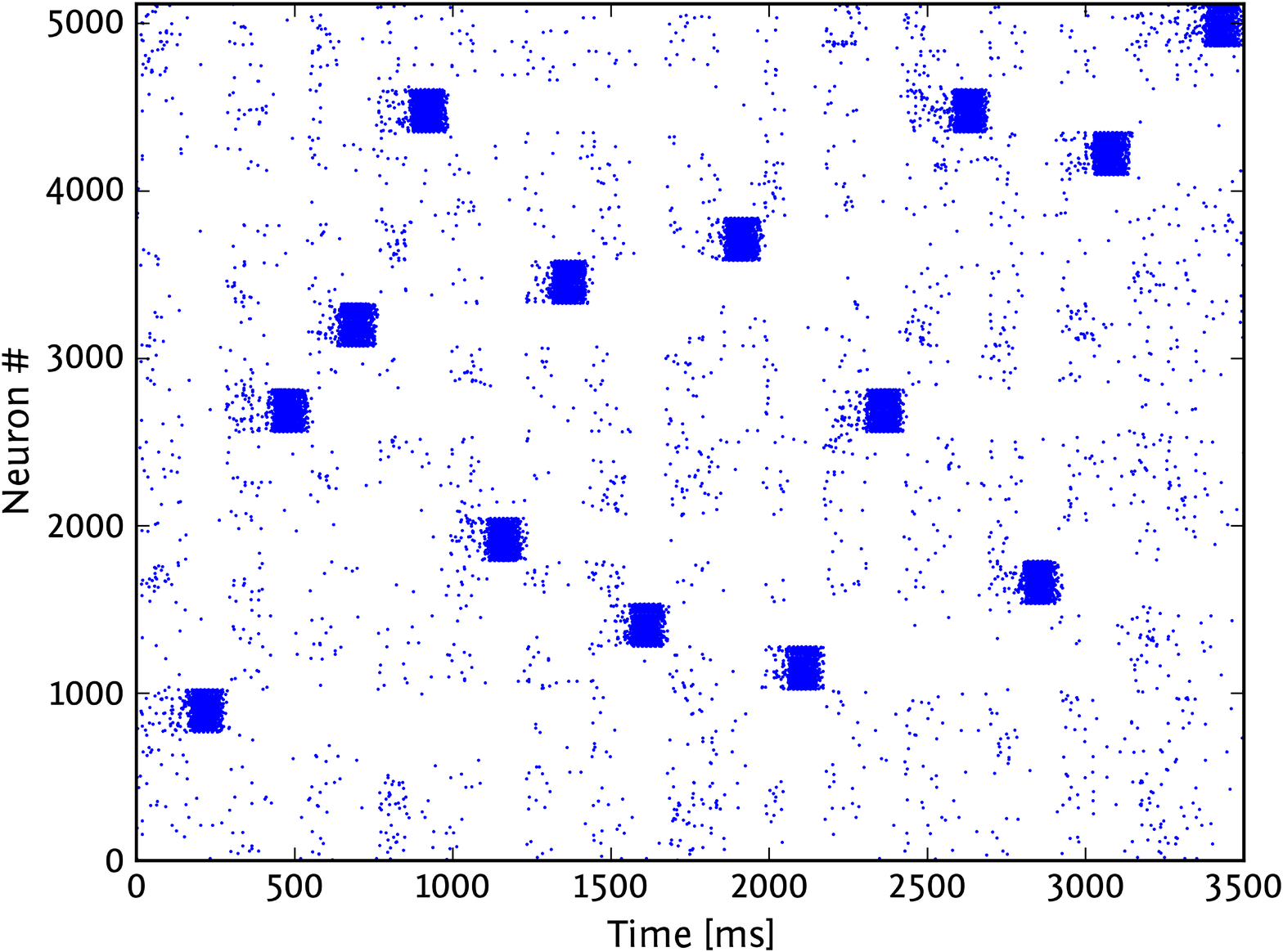}}
	\caption{\label{figure:kthscalingexamples} L2/3 network, scaled down to 9 hypercolumns (HC) with 3 attractors and scaled up to 8 hypercolumns with 20 attractors. Note the relatively long dwell times in \subref{figure:kth9x3scaling} compared to short dwell times in \subref{figure:kth8x20scaling}.}
\end{figure}

Figure \ref{figure:kthscalingexamples} shows a scaling example where the number of attractors is varied. 
At first glance, it may seem that the characteristic attractor dynamics are affected, as the average attractor dwell times decrease from about \unit[300]{ms} to under \unit[200]{ms}. 
However, this is only an apparent effect, as the temporary dominance of individual attractors is a result of local fluctuations in the input. 
An increasing number of attractors means there is more competition among them, which in mathematical terms translates to shorter, smaller fluctuations in the input rate, therefore leading to decreasing dwell times. 
When only two attractors are stimulated, the dynamics are not influenced by the total number of attractors in the network, which supports our scaling rules.

\paragraph{Synfire Chain with Feedforward Inhibition}

\begin{figure}[htb]
	\centering
	\subfloat[5 populations with 64 excitatory neurons each]{\label{figure:synfire_scaling_small}\includegraphics[width=.98\columnwidth]{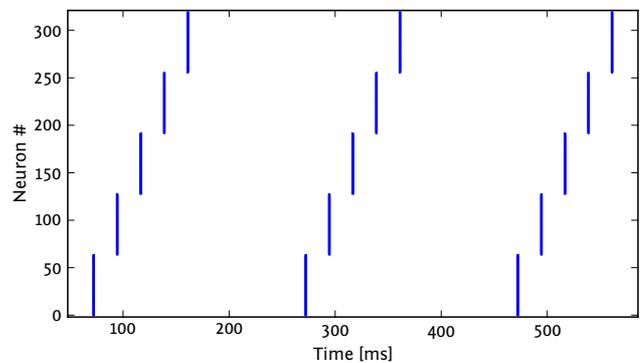}}\\
	\subfloat[32 populations with 200 excitatory neurons each]{\label{figure:synfire_scaling_large}\includegraphics[width=.98\columnwidth]{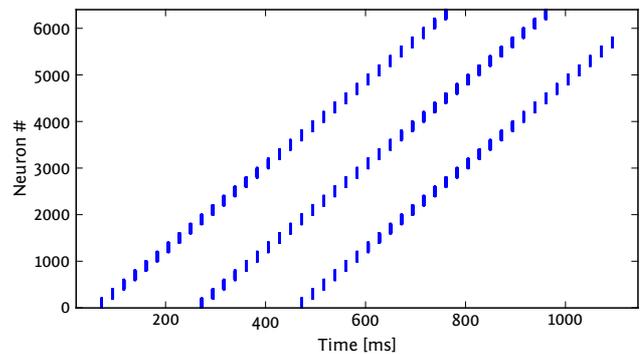}}
	\caption{\label{figure:synfirescalingexamples}Synfire Chain scaling examples.}
\end{figure}

Scaling the Synfire Chain is a comparatively simple task, as there are no feedback or recurrent connections. 
Scaling the number of units does not require any changes in connectivity. 
When the number of neurons per unit is changed, the dynamics can be kept unmodified (synchronous firing within a population) if the number of inputs per neuron remains the same. Therefore, modifying a population size by a factor $\alpha$ simply requires that all connection probabilities are modified by a factor $1/\alpha$. 
Some difficulties may arise when populations become too small, as the binomial connection distribution diverges away from a symmetric Gaussian, favoring a smaller number of afferent connections and leading to activity attenuation and eventually to a break in the pulse transmission \citep{kumar10spiking}. The straightforward remedy is offered by the PyNN class {\tt FixedNumberPreConnector} which guarantees a constant but randomly distributed number of inputs. If populations become too small to accommodate the required number of connections, synaptic weights can be increased to compensate for the synaptic loss. 
The same can be done to cope with synapse loss resulting from the mapping process, as described in Section~\ref{section:distortion_compensation}.
Figure \ref{figure:synfirescalingexamples} shows a scaling example where both size and number of populations are varied.

\subsubsection{Simulating Hardware Imperfections}
\label{section:distortion_mechanisms}

For this study, we have investigated several distortion mechanisms which can be replicated in software simulations and do not require the presence of the actual hardware.

A first limitation of the hardware lies in the fact that axonal and dendritic delays can not be programmed and the intrinsic delays caused by the hardware communication infrastructure are very small when translated to biological real-time. 
This means that, effectively, the hardware can not accurately reproduce the dynamics of networks which rely on delayed transmission of action potentials between two communicating neurons.

Two further distortion sources lie within the synaptic circuits of the HICANN building blocks. 
Since the synaptic weight resolution in a neuromorphic hardware system is limited (see Section~\ref{section:hardware_stdp}), large differences between synaptic weights which are mapped to the same synapse driver may cause significant changes to the synapses at the lower end of the weight spectrum. 
Also, from our experience with the FACETS chip-based systems \citep{bruederle09establishing_hack}, we know that variations in the manufacturing process can lead to a spatial synaptic weight jitter of up to 40\% of the intended value ($\sigma=0.4\mu$, assuming a Gaussian distribution), even after calibration.
This might be fatal for networks which rely on precise tuning of synaptic weights.

Because of the limited bandwidth of the communication layers (see Section~\ref{section:wafer_communication}), synapses may be lost during the biology-to-hardware mapping process. 
Ensuing distortions depend strongly on the network topology and can become arbitrarily high for sufficiently large and densely connected networks.

Additionally, neuron loss can also become a key factor, not necessarily due to hardware limitations (usually, synaptic loss becomes significant long before the number of network neurons exceeds the number of neurons available on the hardware), but as an option to counteract synaptic loss by controlled reduction in the number of neurons. 

Although it does not apply to the three benchmark networks we have studied, the hardware neuron model itself may become a limiting factor, when trying to map models which rely on more complex neurons. 
However, we consider this to be an unlikely case, as the AdEx model has been shown to successfully reproduce a wide variety of spike patterns \citep{brette_05,ALAN09,millner10} and has also proven very successful in emulating biological spiking activity \citep{naud08}.
This is not the case for the FACETS chip-based neuromorphic system (see Section~\ref{section:stage1hardware}), which only offers simple leaky integrate-and-fire neurons.
Section~\ref{section:miniKTH} describes a setup where the missing adaptation mechanism was compensated by tuning several other parameters.

\subsubsection{Analysis and Development of STDP in Hardware}
\label{stdp_methods}

Synaptic plasticity on the highly accelerated FACETS wafer-scale hardware (for a detailed description see Section \ref{section:hardware_stdp}) provides a promising architecture for doing research on learning.
But so far there are no studies about the effect of low weight resolutions and limited update frequencies on the functionality of single synapses and consequently neuronal networks.
In the following, two directions of study will be outlined and one detailed example will be given.

First, the question of a required minimal resolution for synaptic weights and their optimal update frequency is investigated.
However, those two restrictions may be dominated by production process variations that set the upper limit for the functionality of the synapses.
Production process variations cause the supposedly identical circuits for causal and acausal correlation measurement to differ due to variations in their transistors.
This asymmetry limits the accuracy of detecting correlations or in other words causes a correlation offset.
With respect to learning neuronal networks (e.g.\ \citealp{davison06}), we are especially interested in the effects of hardware synapses on their ability to detect synchronous input when embedded into an appropriate architecture.

Secondly, the dynamics of discretized STDP are analyzed based on the assumption that the weight discretization is the most crucial restricting component influencing the dynamics of single synapses and whole networks.
This analysis is carried out with respect to the equilibrium weight distribution that is obtained by evaluating an initial synaptic weight value in sequence. 
Within this sequence of weight evolution, the probability for causal evaluation is equal to the one for acausal evaluation.
Analytical equilibrium distributions \citep{g-q00stable} as well as numerical equilibrium distributions of continuous weights are used as a reference.

Here, we shall discuss in detail one analysis on the effect of low resolution weights within a neuronal network.
In order to isolate the functionality of a single synapse from network effects a simple network is defined (Figure~\ref{stdp_trace}A).
A population of pre-synaptic neurons is connected to a single post-synaptic neuron.
The \emph{Intermediate Gütig STDP model} \citep{guetig03, morrison08_stdp} is used for the construction of the look-up table (see Section \ref{section:hardware_stdp}).
Developing synaptic weights are compared for either correlated or uncorrelated pre-synaptic input.
Correlation within the pre-synaptic population is generated by a multiple interaction process \citep{kuhn03}, whereas in the uncorrelated case the firing pattern of the pre-synaptic neurons are those of Poisson processes.
Results for the effect of discrete weights on this network are presented in Section \ref{stdp_results}.

To avoid expensive changes of the chip layouts, the hardware restrictions are analyzed with preparative software simulations.
Therefore the standard STDP synapse model of the software simulator NEST \citep{scholarpedia_articlenest} was modified by introducing a causal and acausal accumulation circuit, a digital weight value and global weight update controllers.
In the following we will call this the \emph{hardware inspired model}.
As a \emph{reference model} another software synapse model with continuous weight values and continuous weight updates, but with a symmetric nearest-neighbor spike pairing scheme was implemented.

Further analysis with focus on the weight update frequency is in progress.
In the current prototype of the HICANN building block the causal and acausal accumulation circuits will be reset commonly, if a weight update is performed.
Such a common reset distorts the counterbalancing effect of the accumulation circuit receiving less correlations, because the common reset suppresses the circuit to ever elicit a weight update.
Consequently the dominating accumulation circuit, in terms of eliciting weight updates, drives all synaptic weights to its corresponding boundary value.
For future improvements, the performance gained by adding a second reset line to reset both accumulation circuits independently will be compared to the performance gain due to a more detailed readout of these circuits.
Details about these additional studies will be presented in a publication that is in preparation.

\section{Results}
\label{section:results}

In the following, a summary of results is provided, all of which have been acquired by means of the workflow described in Section~\ref{section:methods}.
The presented data demonstrate the functionality of both the PyNN-to-hardware mapping framework and the virtual wafer-scale hardware system, as the applied benchmark models are shown to exhibit the expected functionality when processed by this environment.
Examples of mapping-specific distortion analyses based on reference software simulations are provided and discussed.
The effect of discretized synaptic weights, as implemented in the FACETS wafer-scale hardware, is analyzed in the context of an STDP benchmark.
Scalability questions regarding the graph-based data structure for the mapping process are considered on the basis of experimental data.
Furthermore, we present first results of a successful application of the presented AdEx neuron circuit calibration scheme acquired with a HICANN prototype.

\subsection{Benchmark Results}
\label{section:benchmark_results}

The benchmark models and their target output descriptions described in Section~\ref{section:benchmarks} represent an essential tool to test and verify the workflow presented in this article on a high, functional level.
This is important especially in the context of studies on neural network dynamics, where the identification of erroneous components from the analysis of spatiotemporal spike patterns can be very difficult due to a lack of insight and intuition in the field of neural information processing principles.

\subsubsection{Distortion Mechanisms and Compensation Strategies Based on Software Simulations}
\label{section:distortion_compensation}

Even with a virtual version of the hardware system being available, software simulations remain a powerful analysis tool, especially since they offer access to the full range of dynamic variables of the modeled neural network, some of which may be inaccessible on the virtual hardware. In the following, we will demonstrate the effects of different distortion mechanisms via software simulations and propose several methods to either circumvent or counteract them. These methods are chosen such as to lie within the possibilities of the hardware system.

\paragraph{Layer 2/3 Attractor Memory}

The functionality of the L2/3 network is best determined by a combined analysis of spike and voltage data. 
While a visual inspection of a raster plot of all neurons usually provides a good basis for evaluation, a more thorough investigation of UP/DOWN-state statistics requires the analysis of voltages from a relatively large number of individual cells. 
Both the extraction of the full spike data and of multiple voltage traces are not possible on the hardware, making the use of software simulations indispensable for a proper evaluation of the effects of mapping distortions.

\begin{figure}[htbp]
    \centering
    \includegraphics[width=\columnwidth]{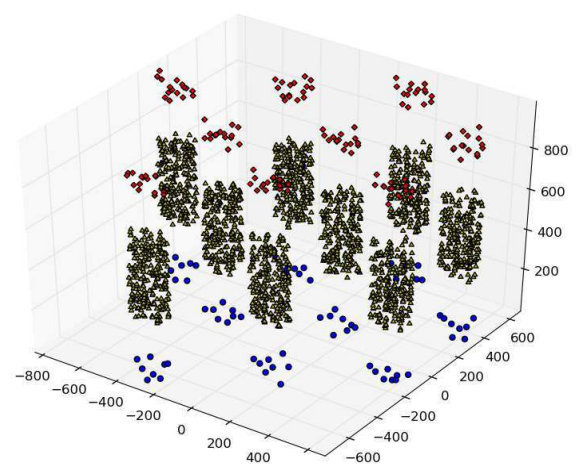}
    \caption{Geometry of the L2/3 Attractor Memory model, the unit on the axes is $\upmu$m.}
    \label{figure:kthgeometry}
\end{figure}

In order to replicate a biologically plausible pattern of axonal and dendritic delays, we have implemented a network geometry as exemplified in Figure~\ref{figure:kthgeometry}, with distance-dependent delays. 
When setting all delays to zero, we have observed no significant changes in the network dynamics. 
This is not unexpected, as this model relies more on firing rates rather than precise spike timing in order for particular attractors to become activated.

\begin{figure}[htb]
    \centering
    \subfloat[Spontaneous activity at 25\% spatial jitter]{\label{figure:kth9x3jitter1}\includegraphics[width=\columnwidth]{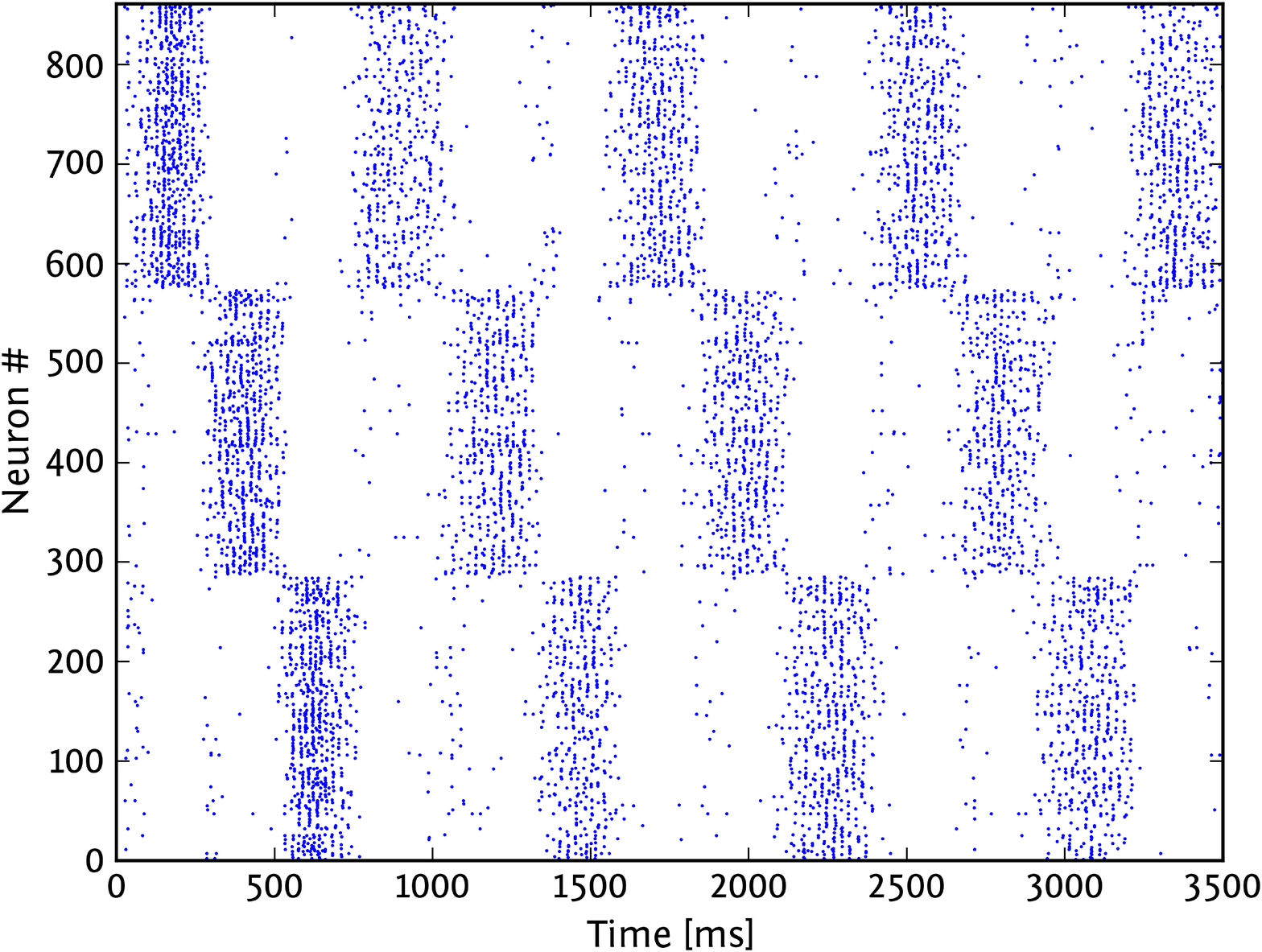}}\\
    \subfloat[L4 activation at 50\% spatial jitter]{\includegraphics[width=\columnwidth]{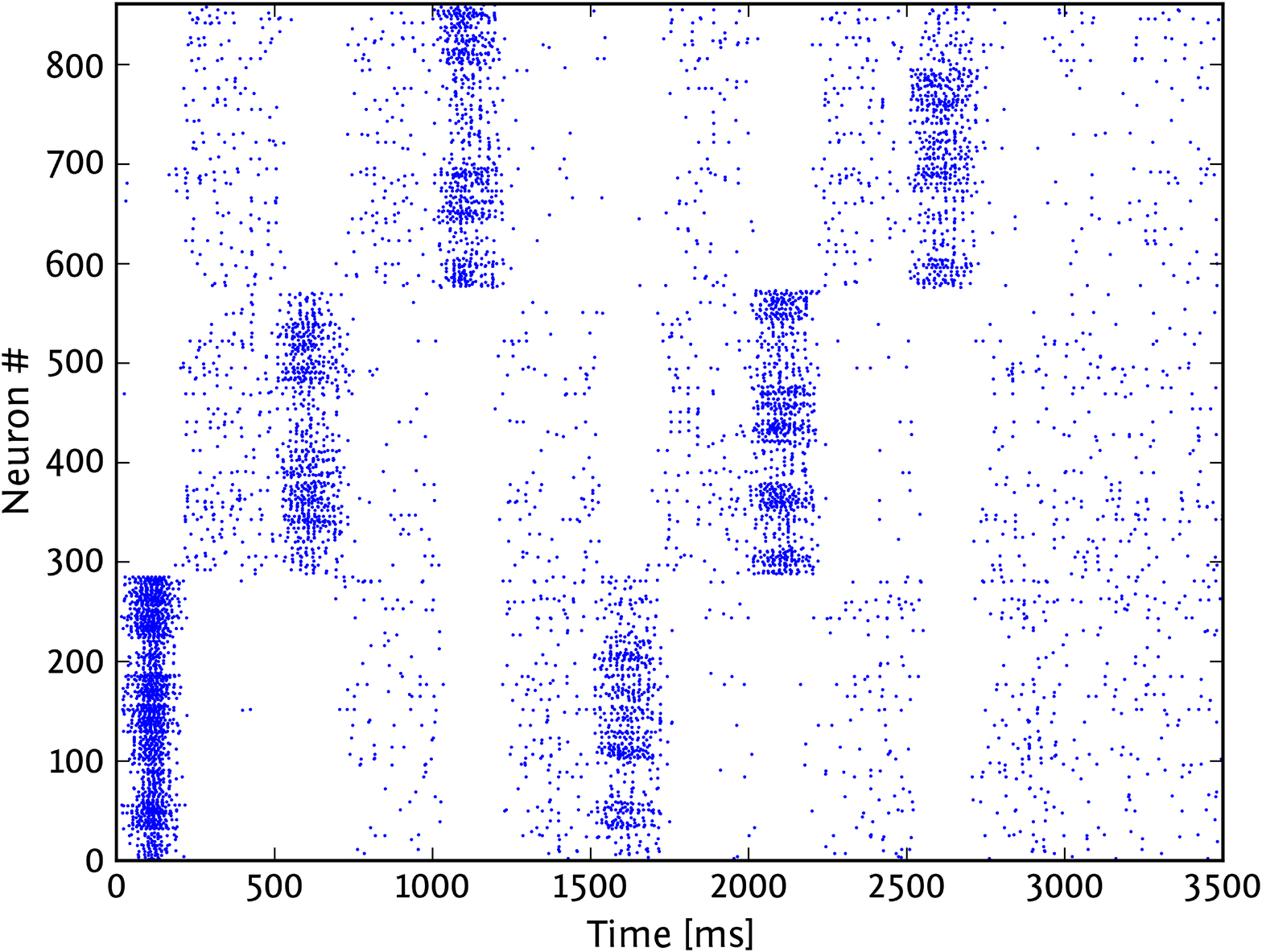}}
    \caption{\label{figure:kthjitterstuff}Effects of spatial weight jitter on a L2/3 Attractor Memory network with 9 hypercolumns and 3 attractors.}
\end{figure}

Depending on the amount of spatial synaptic weight jitter, the network shows varying levels of tolerance. For values up to 25\%, the dynamics only suffer minor changes. At 50\% jitter, spontaneous activation is completely suppressed, but activation via input from L4 remains functional, exhibiting the same phenomena of pattern completion and rivalry as seen in the undistorted case (see Figure \ref{figure:kthjitterstuff}).

\begin{figure}[htbp]
    \centering
    \subfloat[Raster plot, 40\% synaptic loss]{\includegraphics[width=.99\columnwidth]{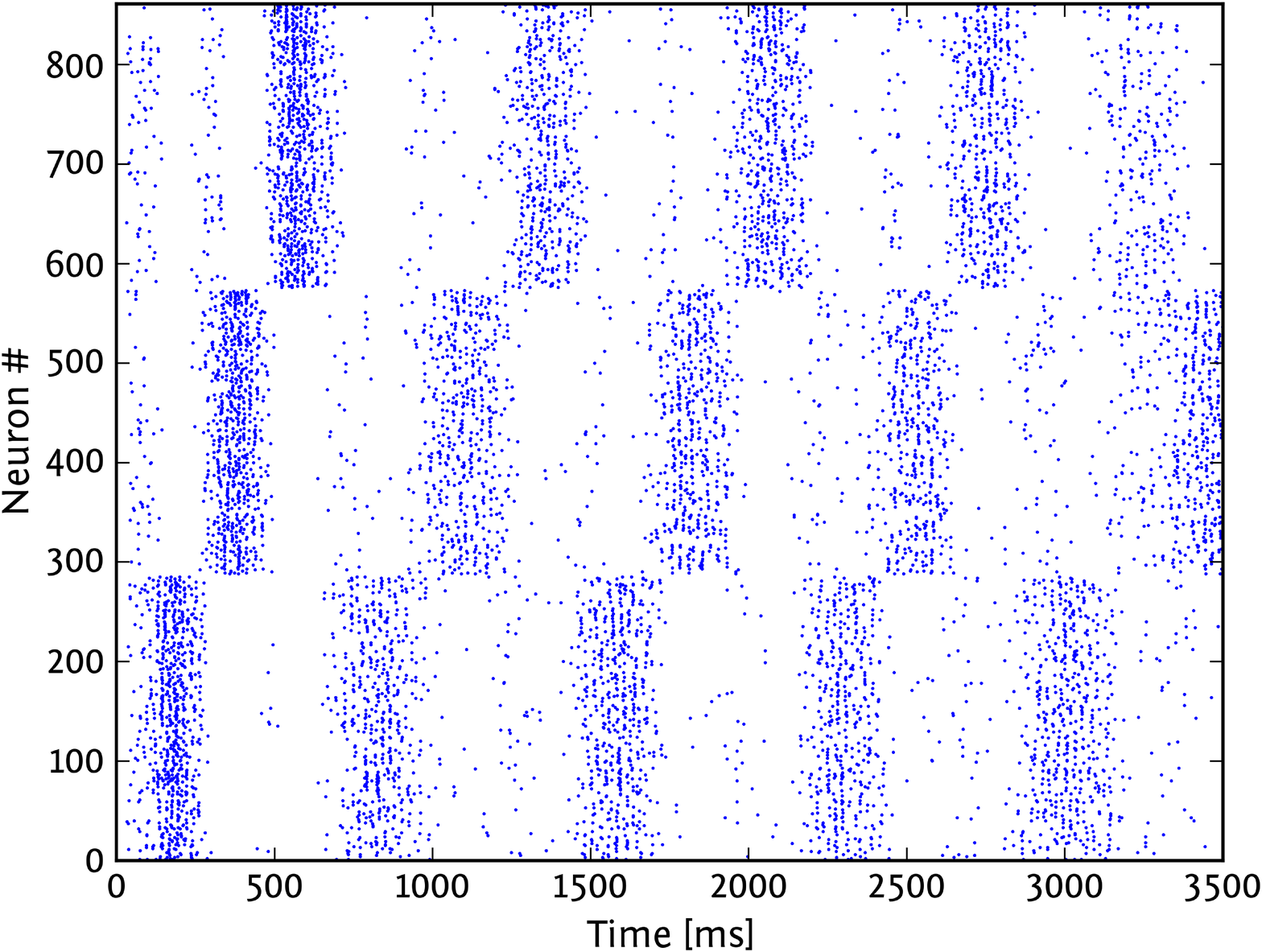}}\\
    \subfloat[Voltage star plot, 40\% synaptic loss]{\includegraphics[width=.99\columnwidth]{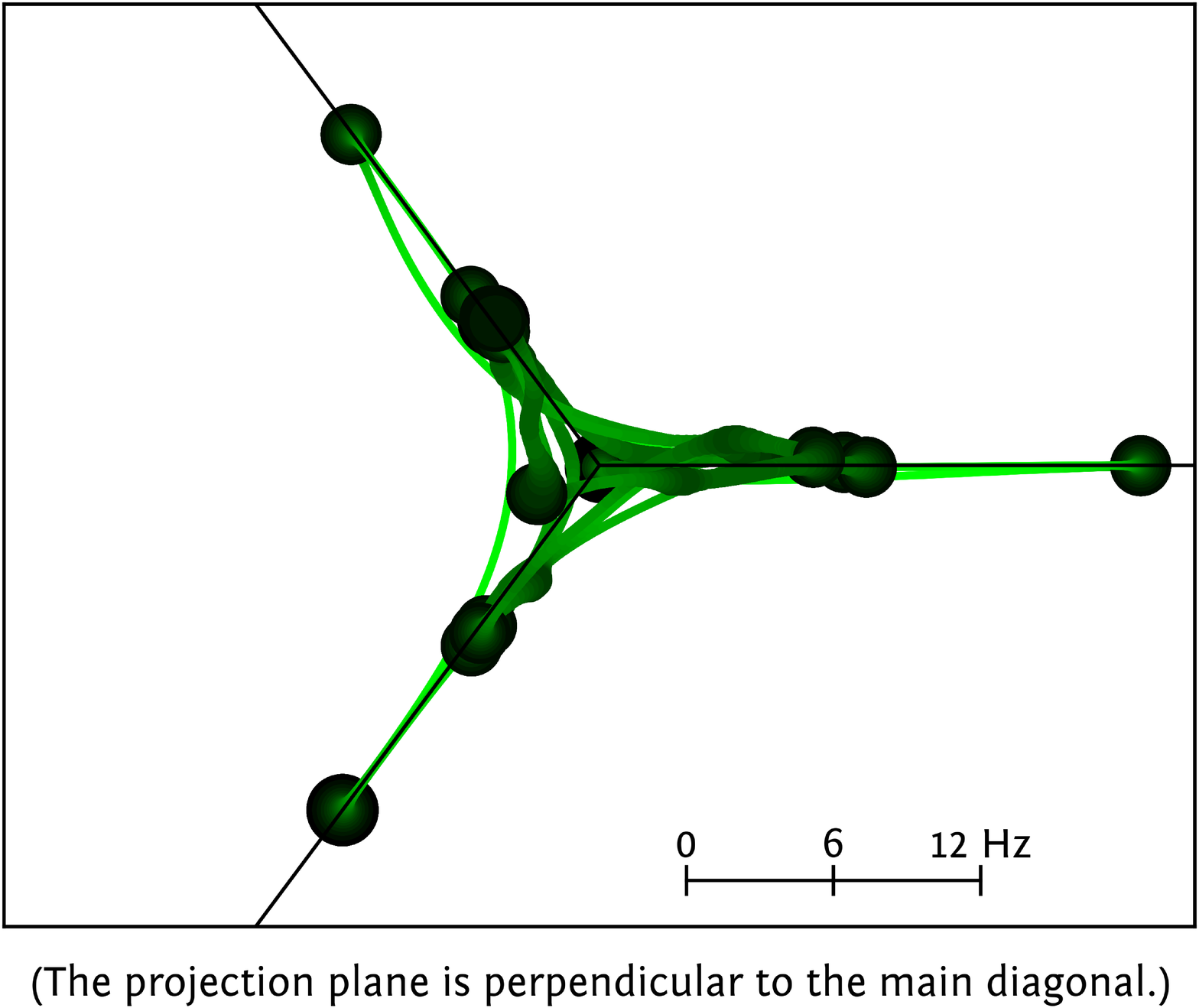}}
    \caption{Synaptic loss tolerance of an L2/3 Attractor Memory network. Synaptic loss was assumed homogeneous over all populations.}
  \label{figure:kth9x3synloss}
\end{figure}

Because of its intricate connectivity, which spans both local and global scales, the Layer 2/3 Attractor Memory network was expected to be quite sensitive to synaptic loss. 
Indeed, if the synapse loss is localized to certain attractors, they become either inactive (for excitatory synapse loss) or dominant (for inhibitory synapse loss). 
However, if synapse loss is spread homogeneously over all populations, the network becomes remarkably resilient, tolerating values as high as 40\% (see Figure \ref{figure:kth9x3synloss}).

\begin{figure}[htbp]
    \centering
    \includegraphics[width=.9\columnwidth]{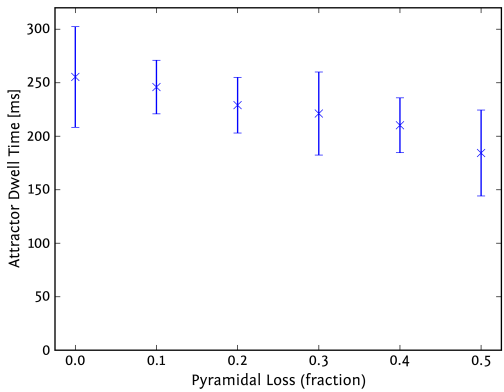}
    \caption{Attractor dwell times versus neuron loss in the L2/3 Attractor Memory network.}
  \label{figure:kthneuronloss}
\end{figure}

In contrast to synaptic loss, the loss of pyramidal neurons (which make up about 87\% of the network) has only little effect on the network dynamics, even up to values as high as 50\%, regardless of the number of minicolumns or hypercolumns present (see Figure \ref{figure:kthneuronloss}). 
It is, for example, possible to have a functioning network with as little as 12 pyramidal cells per minicolumn. This circumstance has major consequences for synapse loss compensation.

When synapse loss increases beyond a certain limit, intra-attractor excitation and inter-attractor inhibition become too weak for attractor dynamics to emerge. The total number of synapses scales linearly with the total number of neurons (when network scaling conserves the afferent fan-in, as described in Section~\ref{subsubsection:Network_Scaling}), so reducing the neuron count represents a straightforward way of circumventing synapse loss. This can be achieved by reducing the number of attractors (which may, however, not always be desirable) or by reducing the number of neurons per attractor by decreasing either the total number of hypercolumns or the neuron count per minicolumn.

Elimination of pyramidal neurons (without re-scaling the fan-in) is a much more efficient method in terms of synapse number reduction, as the total synapse count has an approximately quadratic dependence on the number of neurons per minicolumn. Since, for this particular network model, attractor dynamics are largely insensitive to pyramidal cell elimination, as described above, this becomes a method of choice when dealing with harsh bandwidth limitations.

\begin{figure}[htbp]
    \centering
    \subfloat[60\% synapse loss, uncompensated]{\label{figure:kth9x3synloss60}\includegraphics[width=\columnwidth]{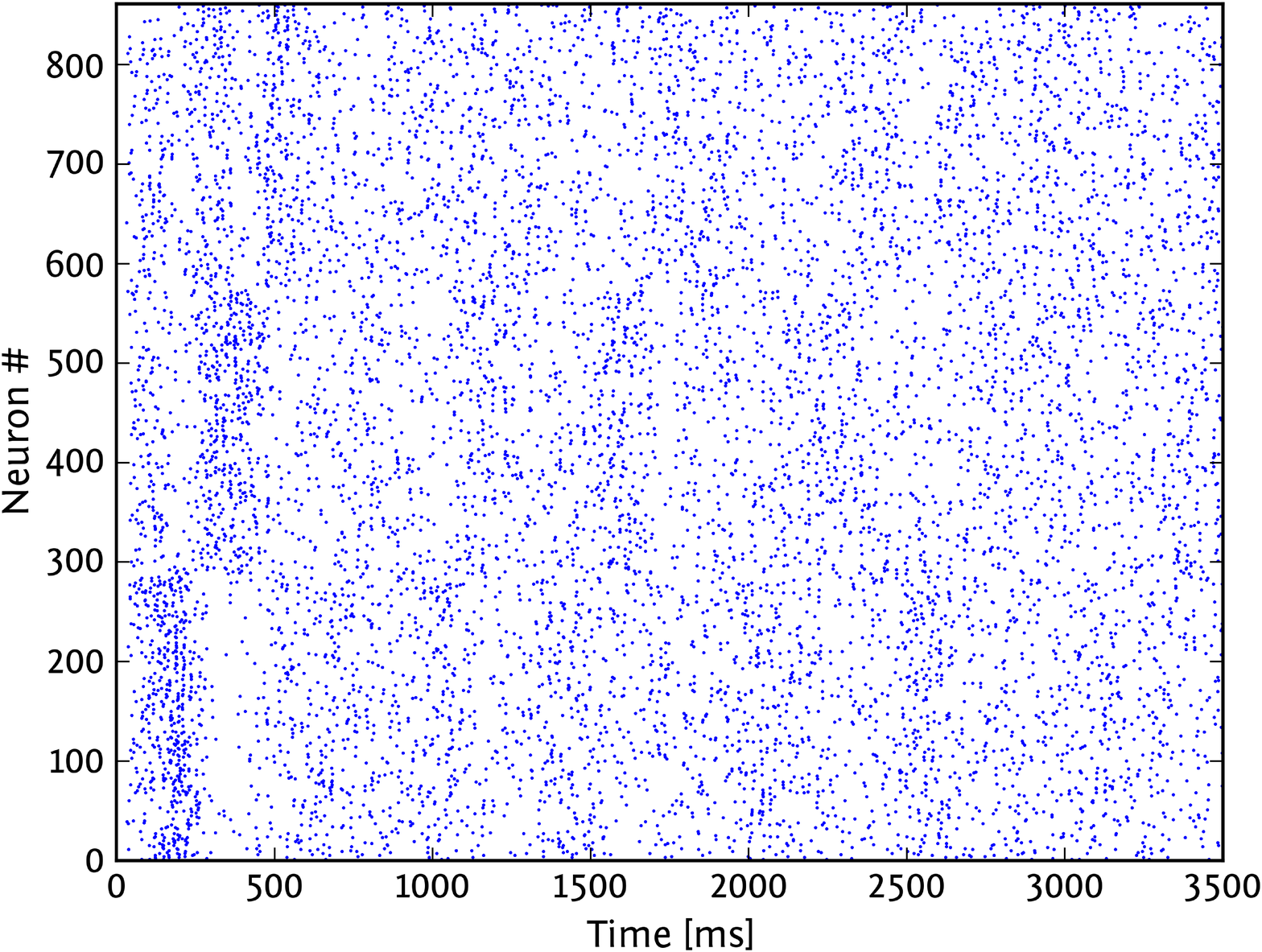}}\\
    \subfloat[60\% synapse loss, compensation by modified weights]{\label{figure:kth9x3synloss60wcomp}\includegraphics[width=\columnwidth]{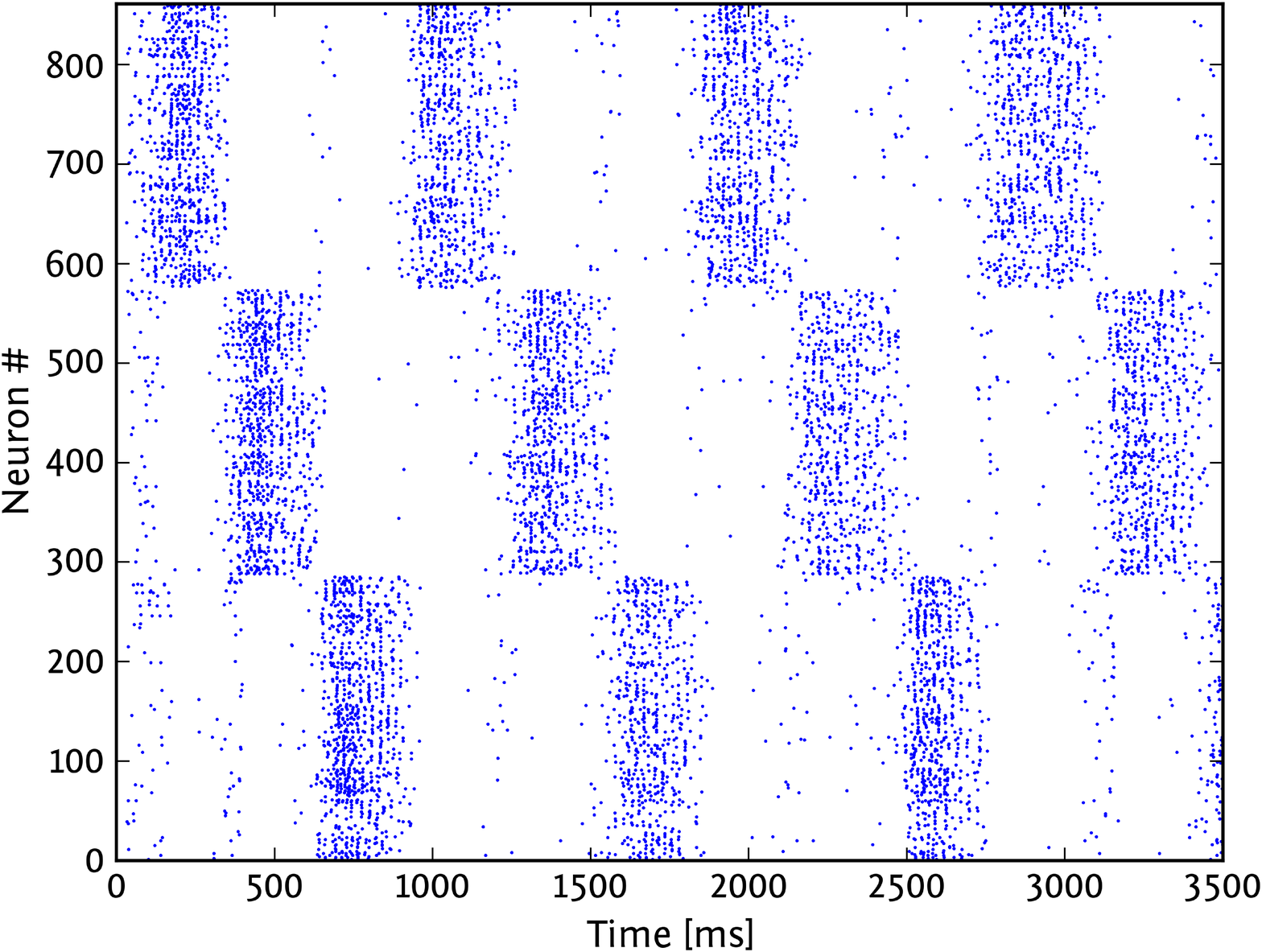}}
    \caption{\label{figure:kthsynlosscomp}\subref{figure:kth9x3synloss60} High synaptic loss destroys attractor dynamics. Several methods for compensating or counteracting this effect are presented in the text above. \subref{figure:kth9x3synloss60wcomp} shows the results of compensation by modified synaptic weights.
}
\end{figure}

Especially in cases where synapse loss is relatively small and inhomogeneous, afferent synaptic input can be restored by increasing the corresponding synaptic weights (see Figure \ref{figure:kthsynlosscomp}). While it is always possible to hereby establish the required average firing rates of individual populations, this compensation mechanism needs to be used cautiously, as it can influence spike train statistics and neuron input-output curves.

\paragraph{Synfire Chain with Feedforward Inhibition}

\begin{figure}
    \centering
    \subfloat[Synfire Chain with delayed spike propagation]{\label{figure:synfiredelay}\includegraphics[width=\columnwidth]{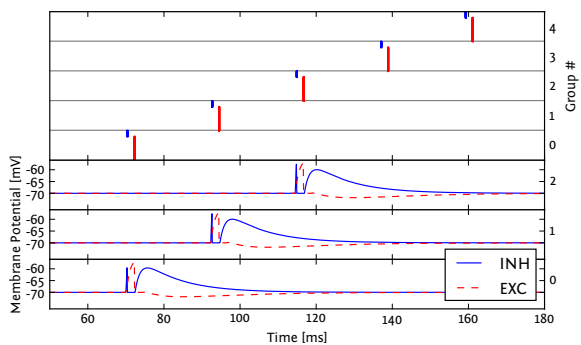}}\\
    \subfloat[Synfire Chain with modified synapses, no delays]{\label{figure:synfirenodelaycompensated}\includegraphics[width=\columnwidth]{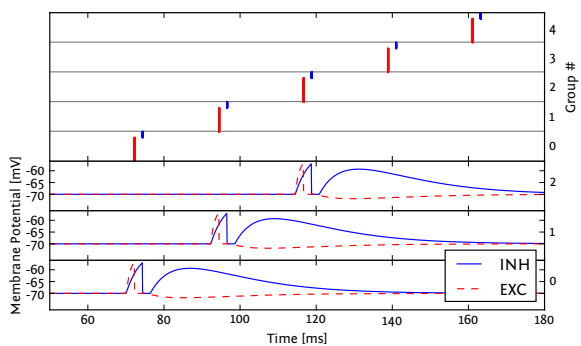}}
    \caption{\label{figure:synfiredelaycompensation}\subref{figure:synfiredelay} Delayed spike propagation is essential in the original Synfire Chain model.\subref{figure:synfirenodelaycompensated} By modifying synaptic parameters (see text for details), effective afferent spike times can be reproduced without propagation delays.}
\end{figure}

The Synfire Chain model presented in Section~\ref{section:synfire_model} relies heavily on delayed transmission of action potentials between inhibitory and excitatory populations. 
Eliminating these delays causes afferent EPSPs and IPSPs to overlap, possibly leading to suppression of efferent spikes \citep{kremkow10gating}. 
This makes a direct mapping of the model to the FACETS wafer-scale hardware impossible. 
However, as the hardware offers the possibility of tuning synaptic time constants of individual neurons, it is possible to compensate for missing delays by adjusting the rising flank of EPSPs for the inhibitory neurons. 
This can be achieved by increasing the corresponding synaptic time constants and decreasing the corresponding synaptic weights simultaneously (see Figure~\ref{figure:synfiredelaycompensation}).

\begin{figure}
    \centering
    \subfloat[Impact of spatial weight jitter on signal propagation]{
        \label{figure:synfirejitter}
        \includegraphics[width=.95\columnwidth]{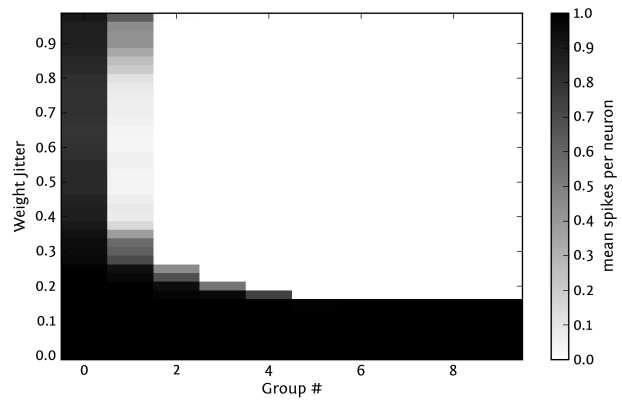}
    }\\
    \subfloat[Impact of synapse loss on signal propagation]{
        \label{figure:synfiresynloss}
        \includegraphics[width=.95\columnwidth]{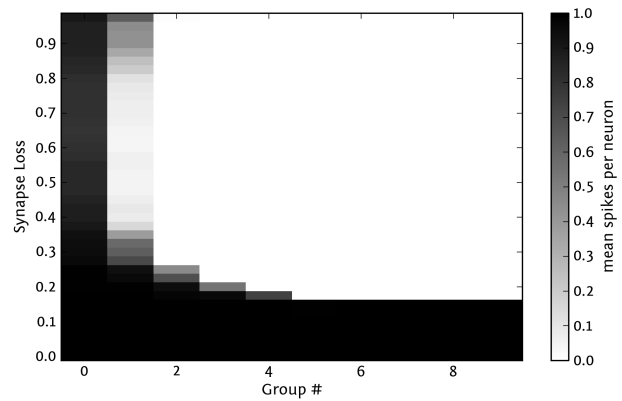}
    }
    \caption{\subref{figure:synfirejitter} Sufficiently high spatial weight jitter causes a breakdown of signal propagation along the Synfire Chain. \subref{figure:synfirenodelaycompensated} Synapse loss is even more critical, completely attenuating the signal after only few iterations.}
\end{figure}

Spatial synaptic weight jitter may effectively cancel itself out for large numbers of synapses, but can lead to skewed afferent input, especially in smaller networks. 
Depending on the amount of spatial jitter (variance of the underlying Gaussian, see Section~\ref{section:distortion_mechanisms}), this might lead to individual excitatory neurons not firing, as a consequence of a too low average afferent weight. 
This causes a chain reaction which leads to an increasing number of silent neurons for every subsequent population in the Synfire Chain, ultimately causing the activity to die out (see Figure~\ref{figure:synfirejitter}).

\begin{figure}[htb]
    \centering
    \includegraphics[width=\columnwidth]{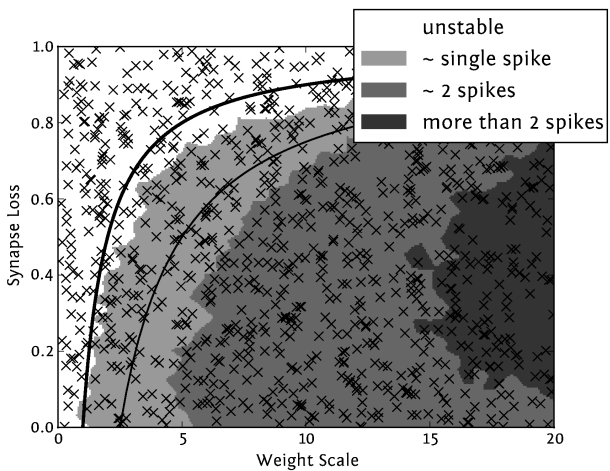}
    \caption{Synapse loss blocks the propagation of the signal along the synfire chain (white zone). A synapse loss probability of $p$ can be compensated very efficiently by scaling the weights by a factor of $\alpha/(1-p)$ (thick black line: $\alpha=1$, thick black line: $\alpha=2.5$). Note that the thin black line stays within the light gray area that denotes a stable propagation with one spike per neuron. Increasing the synaptic weights even more, effectively overcompensating the synapse loss, results in multiple spikes being fired by the excitatory neurons (dark gray and black zone). The total number of spikes per burst is limited by the refractory period and the time until the first inhibitory spike, meaning that the signal does not broaden indefinitely.}
  \label{figure:synfiresynlosscompensation}
\end{figure}

Synapse loss has qualitatively the same effect, only manifests itself much stronger, as it is not symmetrically distributed around zero. 
Even relatively low values of around 2\% completely suppress the propagation of the signal after only few iterations (see Figure~\ref{figure:synfiresynloss}). 
Both distortion mechanisms can be effectively compensated by increasing excitatory synaptic weights (see Figure~\ref{figure:synfiresynlosscompensation}). 
Since all excitatory neurons within a population spike only once, simultaneously, modification of synaptic weights does not affect spike train statistics.

The other obvious way of compensating synapse loss is by decreasing the overall network size, as described for the L2/3 Attractor Memory network. 
This can be achieved by decreasing either the number of populations or their size. Since the mapping algorithm tends to cluster neurons from the same population together on neighboring HICANNs, reducing population sizes is more efficient for reducing the required communication bandwidth.

\subsubsection{Analysis Based on Virtual Hardware}
The benchmark experiments were simulated with the virtual hardware, thereby verifying the functionality of the whole software workflow and the general capability of the system to serve as a neural modeling platform.

Before applying the benchmark experiments (see Section \ref{section:benchmarks}), we determined the maximum reachable bandwidth of the L2 links (FPGA-to-DNC and DNC-to-FPGA) with the aid of the virtual hardware.
We have found that -- despite the Poisson distribution of spiking activity -- the achieved bandwidth corresponds to the one theoretically expected from the data link speed and pulse packet sizes.

The gathered results were used to enhance the routing of L2 pulse events to the wafer (see Section \ref{section:routing}), which distributes external spike sources over all available L2 links, such that the bandwidth provided by a given hardware setup is fully exploited.
The application of this is crucial, when it comes to the realization of network models with either high input rates as the Layer 2/3 attractor memory model,
or highly correlated input as for the Synfire Chain, where hundreds of spikes need to be delivered within a very small time window.
Having these limitations in mind, one can choose the size of the hardware setup properly
before actually mapping neural experiments onto the FACETS wafer-scale hardware, in a way that all requirements are considered in terms of spatial and temporal issues (i.e.\ neuron \slash{} synapse and bandwidth resources).

\begin{figure}[htbp]
\centering
    \includegraphics[width=\columnwidth]{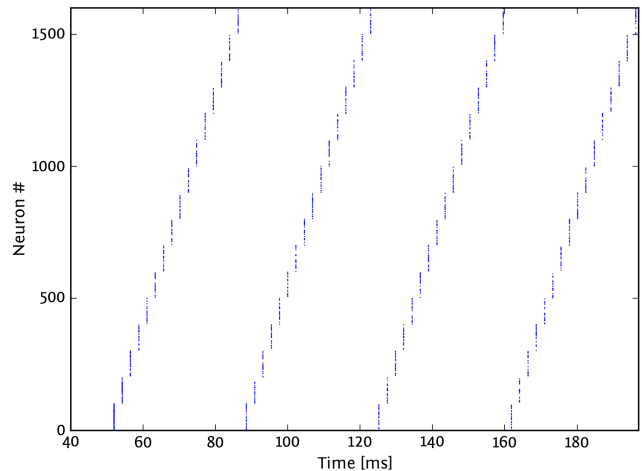}
  \caption{\label{synfire_vhw}
Synfire Chain with 16 groups connected to a loop, simulated on the virtual hardware.
    }
\end{figure}

\paragraph{Synfire Chain}
The Synfire Chain model with feedforward inhibition was successfully run on the virtual FACETS wafer-scale hardware. The stable propagation of pulse volleys from one group to the next is plotted in Figure \ref{synfire_vhw}.
In this case the network consisted of 16 groups with 100 excitatory and 25 inhibitory each, the groups were connected to form a loop such that the activity would be sustained indefinitely.
However, this model proved to be very sensitive to distortions: If more than 2 neurons of a group do not fire, the Synfire Chain will die out immediately, because the local inhibition comes up too early and prevents the excitatory cells from spiking.
This happens also due to the lack of synaptic delays in the current implementation of the hardware, as only L1 connections are used for the routing of neural events within the network.
For upscaled versions of this model and a restricted hardware size, where the mapping process yields a synapse loss larger than $5 \%$, the functionality could not be sustained according to the high sensitivity of the used parameters, i.e. the pulse volley only reached the second group.
Nevertheless, we were able to regain the benchmark functionality by compensating the synapse loss through either strengthened weights or downscaled neuron populations (see Section~\ref{section:distortion_compensation}).

\begin{figure}[htbp]
\centering
    \includegraphics[width=.95\columnwidth]{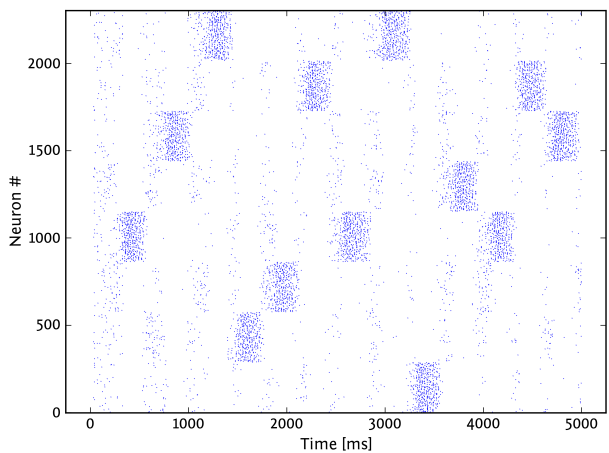}
    \caption{\label{kth_vhw_default} Raster plot of the Layer 2/3 Attractor Memory model simulated on the virtual hardware: firing activity is shown only for pyramidal cells, default size with 9 hypercolumns and 8 attractors.
    }
\end{figure}

\paragraph{Layer 2/3 Attractor Memory Model}
The presented software framework also performed very well when mapping such an elaborated neural architecture like the Layer 2/3 attractor memory model onto the (virtual) FACETS wafer-scale hardware:
Figure \ref{kth_vhw_default} shows the spike output of the default model with 2376 neurons simulated on a virtual wafer snippet containing $8 \times 2$ reticles.
This successful replication of the benchmark's dynamics not only underscores the correct operation of the placing and routing algorithms (see Sections \ref{section:placement} and \ref{section:routing}), but also indicates that the transformation from biological to hardware models (see Section \ref{section:param_translation}) works properly and does not distort the model's behavior, concretely in this example a variety of different short-term plasticity settings could be transferred to shared hardware configurations.

\begin{figure}[htbp]
    \centering
    \subfloat[NEURON]{\label{figure:up_down_neuron}\includegraphics[width=.95\columnwidth]{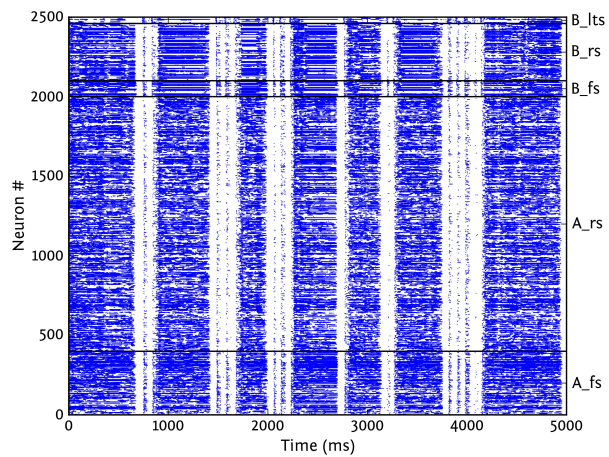}}\\
    \subfloat[Virtual Hardware]{\label{figure:up_down_vhw}\includegraphics[width=.95\columnwidth]{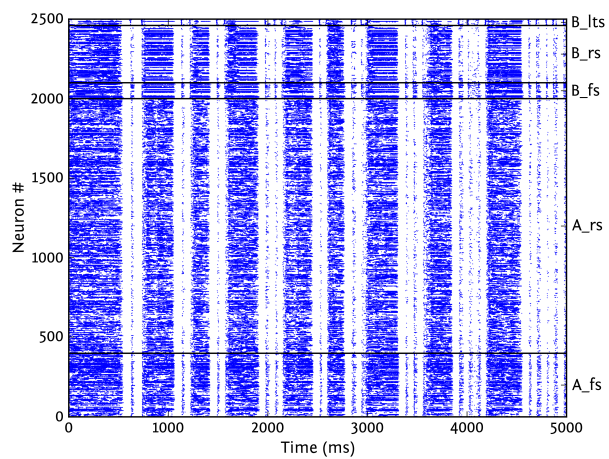}}
    \caption{\label{up_down_raster_plot} Raster plot of the two-layer cortical network exhibiting Up and Down states simulated with NEURON \subref{figure:up_down_neuron} and on the virtual hardware \subref{figure:up_down_vhw}.
Horizontal lines depict the limits between RS, FS and LTS neurons of layers A and B.
The first cortical layer consists of 2000 cells, the second of 500 cells.  10\% of all cells are initially stimulated to induce asynchronous irregular firing in the whole network. The first layer is per se not self-sustaining, i.e.\ the activity dies out after a while, the second smaller layer is able to sustain its activity due to a large number of LTS cells. The sparse connectivity between the two layers assures that the activity in the first is reactivated by excitatory input from the second layer.
}
\end{figure}

\paragraph{Self-Sustained AI States}
The cortical network with self-sustaining AI states was also successfully realized on the virtual hardware.
The single-layer cortical model was implemented for different sizes and parameter sets, where the model functionality was preserved without distortions.
The two-layer cortical network exhibiting Up and Down states was also realized at the default size with 2500 cells and varying adaptation parameters, see Figure~\ref{up_down_raster_plot} for an exemplary raster plot together with a reference software simulation with NEURON.

\subsection{Cross-Platform Implementation of a Benchmark Model}
\label{section:miniKTH}

As a demonstration of the versatility of the methodological framework discussed in the previous sections, this section will present the implementation of one of our benchmark models on three different back-ends: the software simulator NEST, the FACETS chip-based system and the virtual wafer-scale hardware. 
For this purpose, we have chosen the L2/3 Attractor Memory model, due to its challenging connectivity patterns and the interesting high-level functionality it displays. 
Because of the limited size of the chip back-end, the original model needed to undergo some profound changes, which will be detailed in the following sections.

\subsubsection{FACETS Chip-Based Neuromorphic System}

\begin{figure}[htbp]
    \centering
    \includegraphics[width=.95\columnwidth]{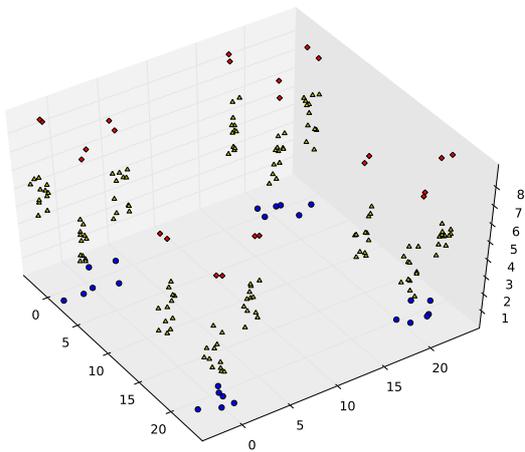}
    \caption{Geometry of the scaled-down L2/3 Attractor Memory network model. Note the greatly reduced number of hypercolumns (HC) and of pyramidal cells per minicolumn (MC) as compared to Figure \ref{figure:kthgeometry}.}
    \label{figure:small_KTH_scaling}
\end{figure}

One ASIC in the current version of the FACETS chip-based system as described in Section~\ref{section:stage1hardware} offers 192 fully interconnectable leaky integrate-and-fire neurons with static synapses. 
Since the original model requires 2376 adapting neurons interconnected through plastic synapses, we had to heavily modify the network configuration in order to keep its functionality.
Reducing the total number of neurons from 2376 to 192 was done following the scaling rules described in \ref{subsubsection:Network_Scaling}. 
In this context, the observation that pyramidal cells can be lost without significantly affecting the dynamics of the network became extremely useful. 
In order to provide relatively long dwell times, we have chosen a setup with only three attractors and four hypercolumns (i.e. four minicolumns per attractor). 
The number of basket cells per hypercolumn was reduced from the original 8 to 6, while the number of pyramidal cells per hypercolumn was reduced from the original 30 to 12. 
The number of RSNP cells per minicolumn remained constant at 2. 
Thus, this setup implements the original model architecture with exactly 192 neurons.
See Figure~\ref{figure:small_KTH_scaling} for a schematic of the resulting architecture.

Due to a lack of neural adaptation and synaptic plasticity (which are both crucial in the original model, as they limit the pyramidal cell UP-state duration), we needed to adapt the neuron parameters (leak conductance, reset, rest, threshold and reversal potentials) and synapse characteristics (weight, decay time constant) in such a way as to retain as much as possible of the original dynamics, on average. 
One additional constraint which needed to be taken into account was the limited range of synaptic weights available on the neuromorphic chip. 
We were able to compensate this, to some extent, by modifying the connection densities among the neuron populations. 

One important consequence is that because the network is unable to adapt, its dynamics change significantly. If one would only remove adaptation and plasticity, without changing other parameters, the first attractor which becomes activated would last indefinitely. Therefore, the removal of these two mechanisms needs to be accompanied by a reduction of intra- and inter-columnar excitation. This in turn causes the network to become much more input driven, which manifests itself in an extreme sensitivity of attractor dwell times towards the momentary input activity. Dwell times become more erratic and even small changes in the average input rate cause attractors to become either dominant or virtually inactive.

Also, due to the limited input bandwidth of the ASIC (for the chosen architecture: 64 channels at about \unit[80]{Hz}), some degree of input correlation was inevitable, as each of the 144 pyramidal cells requires a Poisson stimulation of \unit[300]{Hz}. 
In order to maintain attractor stability, we have chosen to have no overlapping inputs for different attractors (and thus zero correlation, for the Poisson input we have used). 
This, on the other hand, leads to an increased input correlation among pyramidal cells belonging to the same attractor, which, in absence of adaptation, tends to prolong attractor dwell times.
\begin{figure}[htbp]
    \centering
    \subfloat[Poisson input with an overall rate of 750, 1 and 550 Hz]{
        \label{figure:spikey_kth_1}
        \includegraphics[width=0.97\columnwidth]{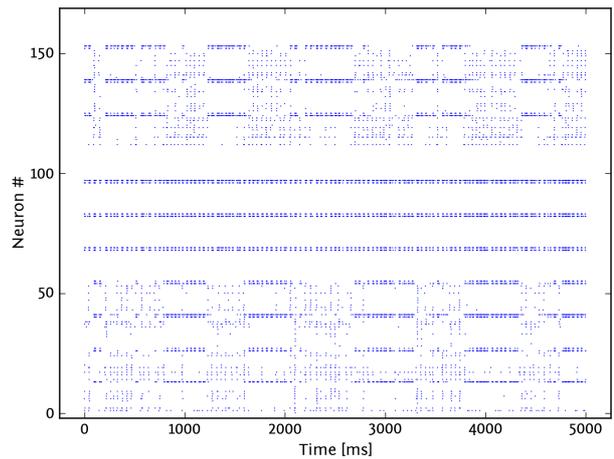}
    }\\
    \subfloat[Poisson input with an overall rate of 750, 700 and 650 Hz]{
        \label{figure:spikey_kth_2}
        \includegraphics[width=0.97\columnwidth]{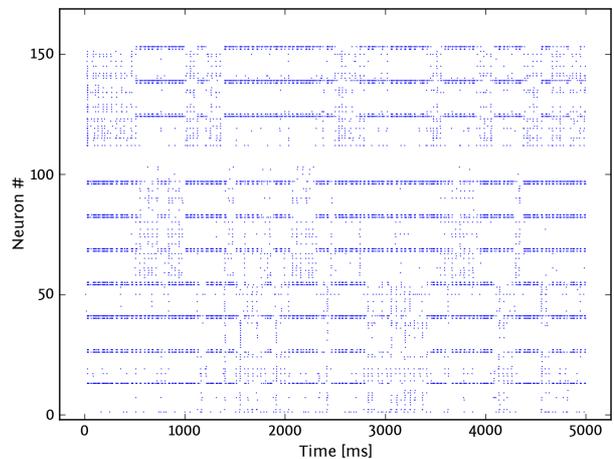}
    }\\
    \subfloat[Poisson input with an overall rate of 700, 800 and 600 Hz]{
        \label{figure:spikey_kth_3}
        \includegraphics[width=0.97\columnwidth]{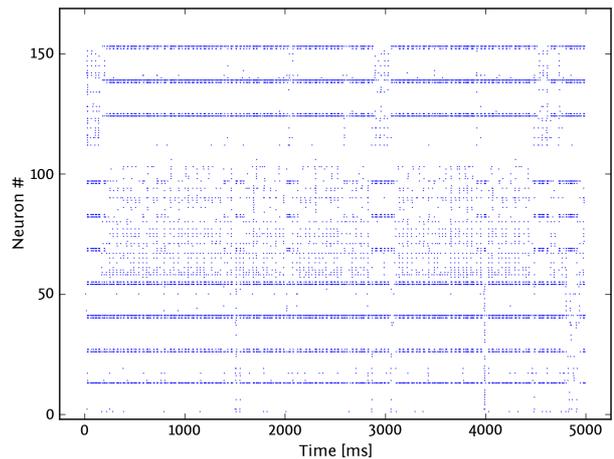}
    }
    \caption{Raster plots of scaled-down L2/3 Attractor Memory networks on the FACETS chip-based neuromorphic system. For explanation see text.}
    \label{figure:spikey_kth}
\end{figure}

Figure~\ref{figure:spikey_kth} shows L2/3 Attractor Memory benchmark results acquired with the FACETS chip-based system:
\subref{figure:spikey_kth_1} Attractors have been excited by Poisson input with an overall rate of 750, 1 and 550 Hz, respectively. 
Note the relatively long dwell times, which are mostly due to high correlations among pyramidal inputs within an attractor. 
The discrepancy in the input activity needed to ensure a balanced activation of attractors 1 and 3 is due to hardware manufacturing fluctuations, which appear to be very complex in nature, often interacting with each other and being highly dependent on the ongoing activity on the chip. 
\subref{figure:spikey_kth_2} Attractors have been excited by Poisson input with an overall rate of 750, 700 and 650 Hz, respectively. 
Note the large fluctuations in attractor dwell time due to lack of adaptation which leads to strongly input-driven dynamics. 
Also note that, in contrast to Figure~\ref{figure:spikey_kth_1}, when attractor 2 becomes active, the input activity of attractor 3 required an increase of 100 Hz in order to achieve balanced activation. 
The most likely cause is capacitive cross-talk between the analog circuits, which varies depending on the throughput rate. 
\subref{figure:spikey_kth_3} Attractors have been excited by Poisson input with an overall rate of 700, 800 and 600 Hz, respectively. 
Note that only a slight increase of the attractor 2 input rate, with respect to the other attractors, results in almost complete dominance of attractor 2. Again, this is due to the lack of adaptive mechanisms.

\subsubsection{Virtual Hardware}

The scaled-down version of the L2/3 Attractor Memory network model was successfully implemented on the \emph{virtual hardware} (see Section \ref{section:virtual_hardware}).
No changes had to be applied to the model in order to realize it on the FACETS wafer-scale virtual hardware, as the HICANN building block implements AdEx-type neurons, which include the dynamics of the leaky integrate-and-fire neurons from the chip-based system.
The scaled-down model passed through the whole mapping process described in Sections \ref{sec:mapping_process}, \ref{section:placement}, \ref{section:routing} and \ref{section:param_translation} and was finally mapped and simulated on a snippet of $2\times2$ reticles of a wafer.

\begin{figure}[htbp]
    \begin{center}
        \subfloat[Poisson input with overall rate of 550, 1 and 550 Hz]{
            \includegraphics[width=0.97\columnwidth]{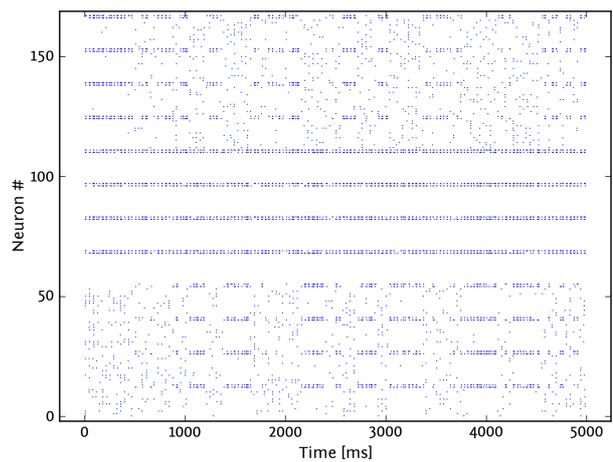}
            \label{figure:vhw_kth_1}
        }\\
        \subfloat[Poisson input with overall rate of 700 Hz]{
            \includegraphics[width=0.97\columnwidth]{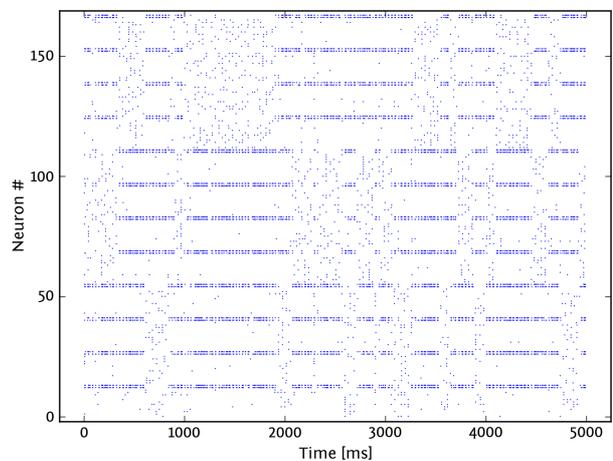}
            \label{figure:vhw_kth_2}
        }\\
        \subfloat[Poisson input with overall rate of 700, 800 and 700 Hz]{
            \includegraphics[width=0.97\columnwidth]{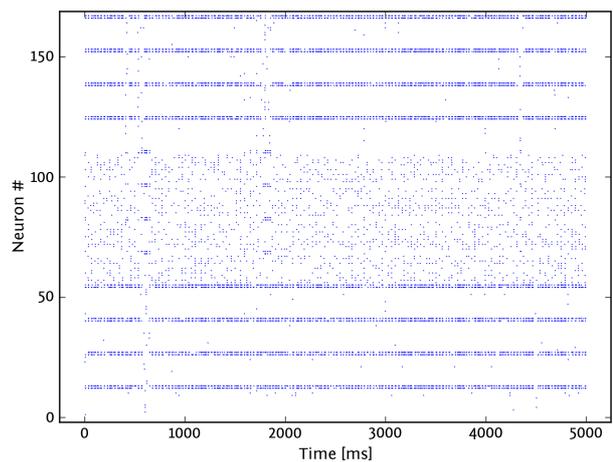}
            \label{figure:vhw_kth_3}
        }
        \caption{Scaled-down L2/3 Attractor Memory network simulated with the \emph{virtual hardware}. Attractors have been excited by Poisson input with different overall rates.}
        \label{figure:vhw_kth_small}
    \end{center}
\end{figure}

\begin{figure}[htbp]
    \centering
    \includegraphics[width=.95\columnwidth]{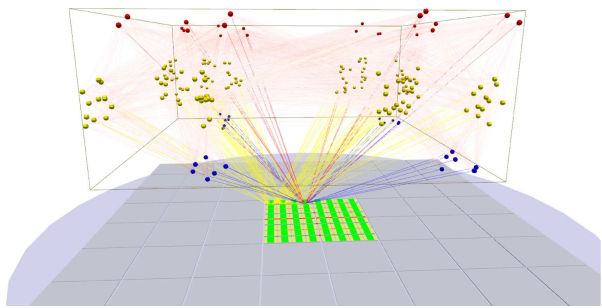}
    \caption{3-D view of the scaled-down L2/3 Attractor Memory network model (see Figure~\ref{figure:small_KTH_scaling}) and its mapping to the wafer generated by the GraViTo software (Section \ref{section:mapping_analysis}).}
    \label{figure:small_KTH_gravito}
\end{figure}

The results of the virtual hardware simulation of the scaled-down L2/3 Attractor Memory network can be seen in Figure \ref{figure:vhw_kth_small}, where the individual attractors were stimulated with different rates.
Depending on the specific stimulation, the network reliably exhibits the same behavior as reference software simulations, which are described in the following section.
Figure \ref{figure:small_KTH_gravito} shows the 3-D visualization of the network model and its mapping onto the wafer with the GraViTo software (see Section~\ref{section:mapping_analysis}).

\subsubsection{NEST Simulator}

The same PyNN script as in the previous sections was eventually used with the software simulator NEST, as a means of comparison to an \textit{ideal} simulation environment. 
By providing identical parameter settings, one can hereby gain a good perspective for gauging the effects of the hardware-inherent fluctuations.

\begin{figure}[htbp]
    \centering
    \subfloat[Poisson input with overall rate of 550, 1 and 550 Hz]{
        \label{figure:nest_kth_1}
        \includegraphics[width=0.97\columnwidth]{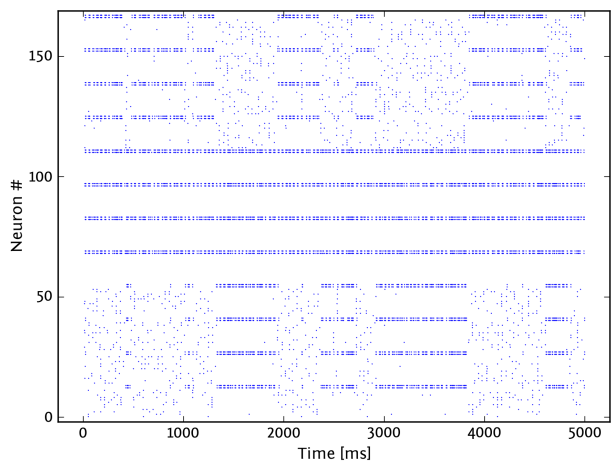}
    }\\
    \subfloat[Poisson input with overall rate of 700 Hz]{
        \label{figure:nest_kth_2}
        \includegraphics[width=0.97\columnwidth]{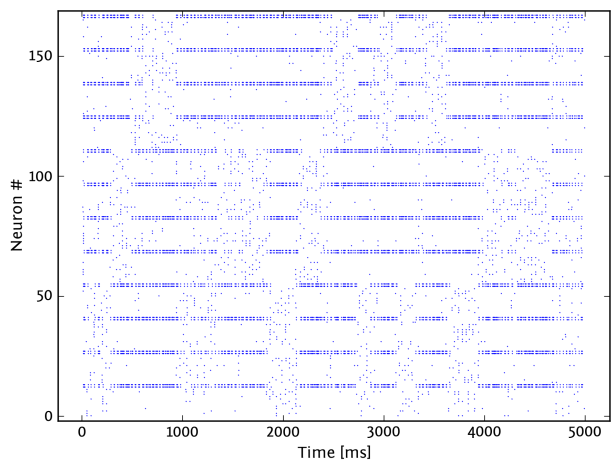}
    }\\
    \subfloat[Poisson input with overall rate of 700, 800 and 700 Hz]{
        \label{figure:nest_kth_3}
        \includegraphics[width=0.97\columnwidth]{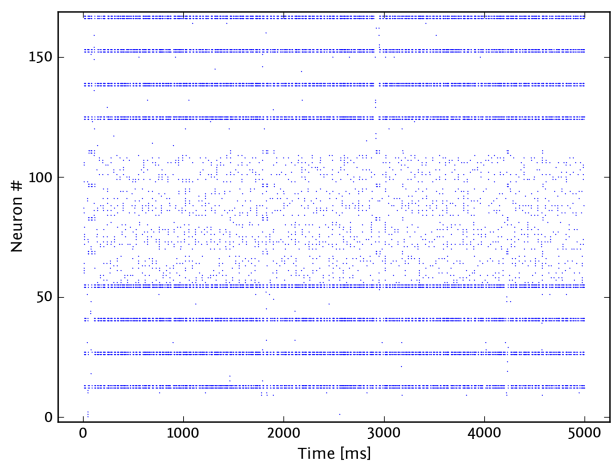}
    }
    \caption{\label{figure:nestkthsmall}Scaled-down L2/3 Attractor Memory network simulated with the NEST software. Attractors have been excited by Poisson input with different overall rates.}
\end{figure}

The results are practically identical to the ones from the virtual hardware, perhaps not surprisingly, as it is not subject to hardware-specific manufacturing process fluctuations (see Figures \ref{figure:vhw_kth_small} and \ref{figure:nestkthsmall}). 

Also, due to its small size, does the network not pose any challenge to the mapping algorithm, making the hardware realization a perfect replica of its software counterpart. 
Still, the successful emulation offers a convincing proof of the efficacy of our mapping work flow.

The chip-based neuromorphic device, on the other hand, is subject to the full range of hardware-specific distortions. 
Nevertheless, the resulting network dynamics agree well with the NEST results, requiring only small adjustments in the input activity. 
These results are expected to greatly improve on the wafer-scale hardware, thanks to the superior architecture of the HICANN units. 
Also, a much more complex neuron model and the availability of both short- and long-term synaptic plasticity mechanisms will make the wafer-scale hardware much more capable of emulating biologically accurate network models.

As a conclusion, we note that the software results are in very good agreement with the ones generated by our hardware back-ends, thus supporting our work flow concept and solidifying the position of our neuromorphic hardware as a universal modeling tool. The particularly appealing feature, especially from a neural modeling perspective, is the seamless transition from software simulation to hardware emulation, which, from the perspective of the PyNN user, is accomplished by modifying a single line of code.

\begin{figure*}[h!]
  \mbox{
    \hspace{0.05\textwidth}
    \mbox{\includegraphics[width=0.9\textwidth]{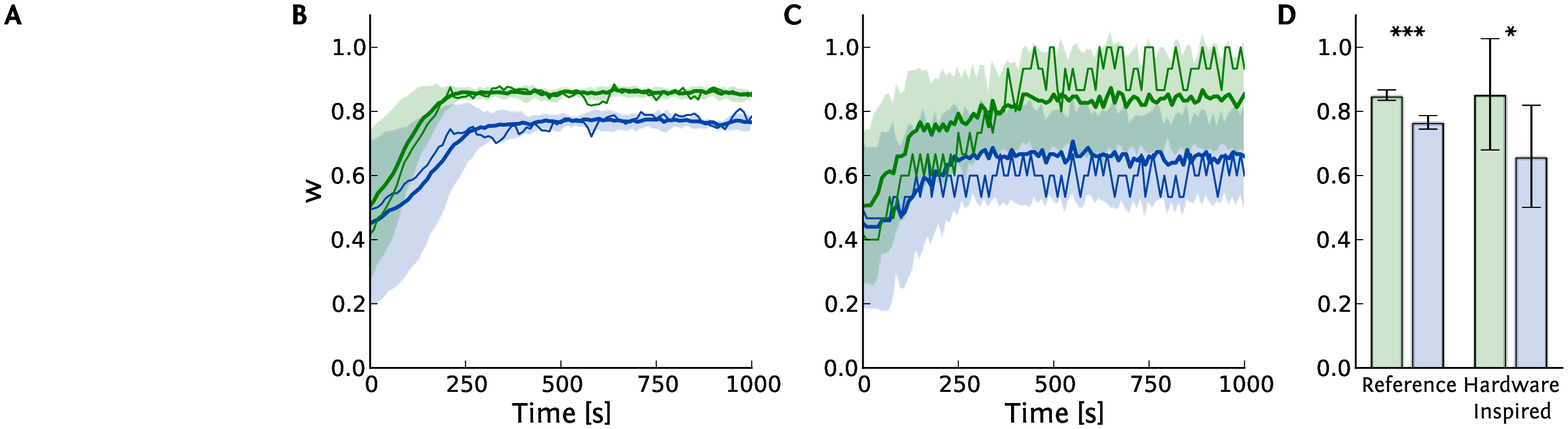}}
    \hspace{-0.899\textwidth}
    \raisebox{0cm}{\mbox{\includegraphics[width=0.135\textwidth]{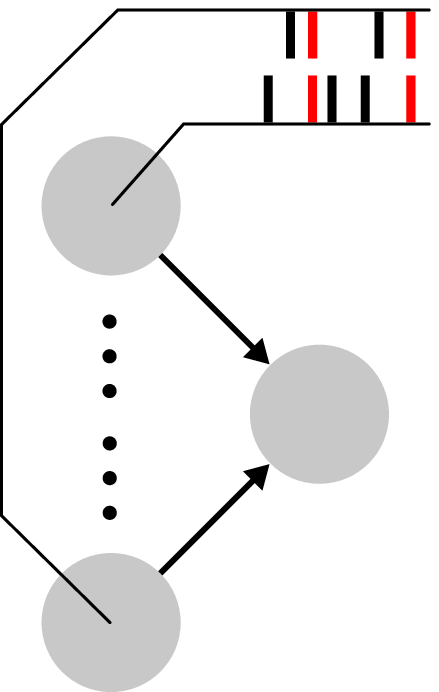}}}
  }
  \caption{\label{stdp_trace}
    {\bf A} STDP evaluation network layout with variable correlation strength $c$.
            Gray circles represent neurons, arrows synapses.
            The spike trains of neurons contain random (black) and correlated (red) spikes.
    {\bf B} Mean (thick line) and standard deviation (shaded area) of reference STDP synapses with correlated (green) and uncorrelated (blue) pre-synaptic neurons.
    Thin lines are single example traces.
    {\bf C} Like B, but with hardware inspired STDP synapses.
    {\bf D} Weight distributions after $\unit[1000]{s}$ for reference ($p<0.001$) and hardware inspired ($p<0.05$) STDP synapses.
  }
\end{figure*}

\subsubsection{Analysis and Development of STDP in Hardware}
\label{stdp_results}

The effects of discrete synaptic weights on networks is analyzed by means of a simple network (Figure \ref{stdp_trace}A).
Ten pre-synaptic neurons are connected to one post-synaptic neuron using both the reference and the hardware inspired STDP synapses as described in Section \ref{stdp_methods}.
In order to analyze effects of discrete weights isolated from other hardware specific constraints the weight update frequency is set equal to the time resolution $h=\unit[0.1]{ms}$ of the software simulator.
The spike rates of the pre-synaptic neurons are adapted in such way that the post-synaptic neuron is firing at about $\unit[10]{Hz}$.
In case of correlated pre-synaptic neurons their correlation coefficient is $c=0.05$ \citep{kuhn03}.
Varying the input spike rates or the correlation coefficient does not change the conceptual outcome.
As the currently implemented hardware synapses have a weight resolution of 4 bits, this resolution is used to test the performance of the hardware.

Figure \ref{stdp_trace} shows the mean weight traces for runs with correlated pre-synaptic neurons as well as for separate runs with uncorrelated pre-synaptic neurons.
In case of hardware inspired STDP synapses (Figure \ref{stdp_trace}C) the standard deviations of the mean weight traces are much larger than those of the reference STDP synapses.
These increased deviations are due to the large weight steps between adjacent discrete weights.
Applying a t-test to the synaptic weight distribution after $\unit[1000]{s}$ shows that the hardware inspired STDP synapses can nevertheless distinguish between uncorrelated and correlated input.
For hardware inspired STDP synapses the probability that the synaptic weights of both populations are separated is $p=0.02$, compared to $p=2 \cdot 10^{-8}$ for the reference STDP synapses.
This ability of distinction determines the ability of detecting synchronous input, which is fundamental for most STDP applications.
For correlation coefficients as low as $c=0.05$ a resolution of 4 bits is still sufficient to detect synchronous pre-synaptic firing.

\subsection{Software Performance}
\label{section:software_performance}

The usability of any hardware modeling platform strongly depends on the time needed for configuration and reprogramming, thus the benchmarks introduced in Section~\ref{section:benchmarks} also serve as tests for \textit{scalability} in terms of time and space.

\begin{figure}[htbp]
    \centering	
    \includegraphics[width=0.8\columnwidth]{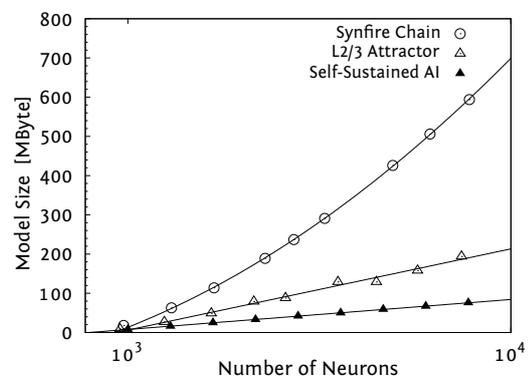}
    \caption{Mapping process scaling in terms of \textit{BioModel} sizes \citep{ehrlich2010anniip}.}
    \label{figure:mapping_scaling}
\end{figure}

Figure~\ref{figure:mapping_scaling} shows that the space consumption for the \textit{BioModel} data grows almost linearly depending on the number of neurons and the synaptic density $\rho_{Syn}$.
Thus, for the given benchmarks, the model sizes for networks with a neuron count $N_{BIO}\leq10^5$ and an approximate average $\rho_{Syn}\leq10\%$ stay within an acceptable limit of \unit[10]{GByte}.
Furthermore, the placement algorithms, in spite of the cubical problem, grow below $O(n^2)$
and as such fulfill the requirement of a reasonable runtime for complex mapping problems \citep{ehrlich2010anniip}.

\subsection{HICANN Prototype Calibration}
\label{section:calibrationresults}

In order to provide an example of its functionality, the calibration framework described in Section~\ref{section:param_translation} has been used to reproduce a biologically relevant tonic spiking neuron on the HICANN prototype. 
First, a reference simulation of a tonic spiking neuron using the AdEx model was created. 
For this simulation, the adaptation and exponential terms were disabled. 
The simulated neuron showed a firing rate of $\unit[53.4]{Hz}$, which due to the speedup factor of the HICANN system corresponds to $\unit[534]{kHz}$ in the hardware domain.

\begin{figure}[htbp]
	\centering
	\subfloat[Reference simulation]{\label{figure:CalibResTop}\includegraphics[width=.95\columnwidth]{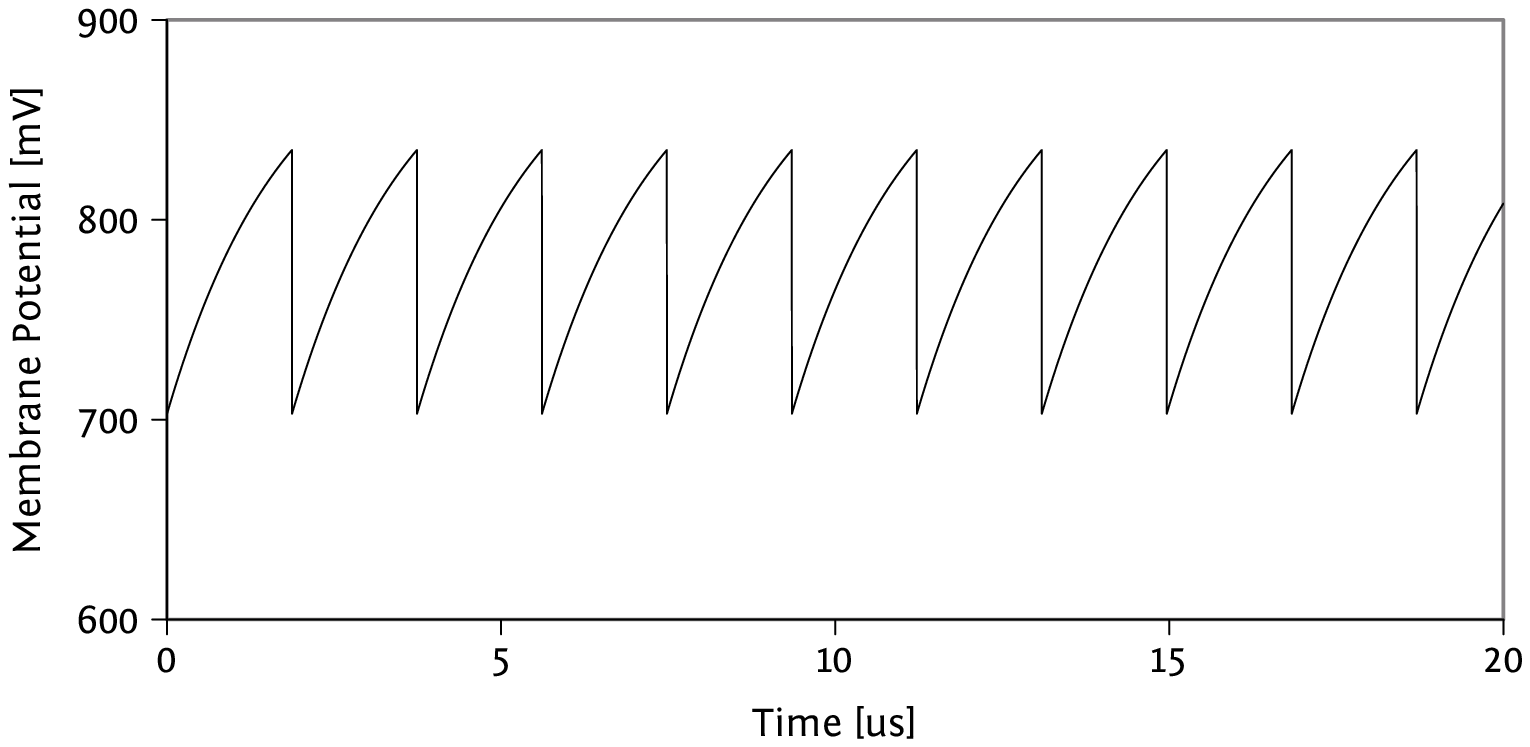}}\\
	\subfloat[Hardware neuron recording]{\label{figure:CalibResBottom}\includegraphics[width=.95\columnwidth]{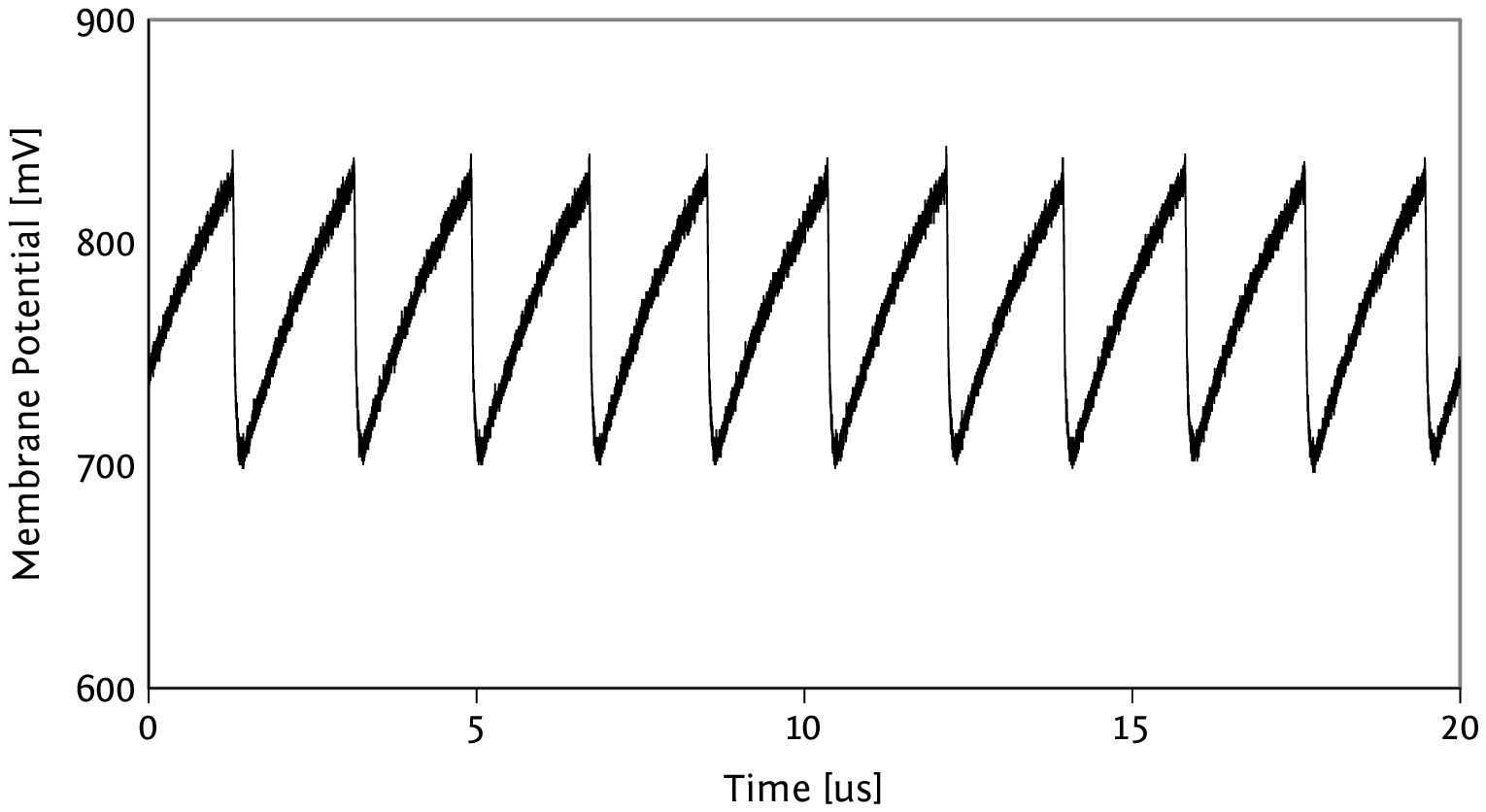}}
	\caption{Comparison between the reference simulation \subref{figure:CalibResTop} and the membrane recording of the hardware neuron after calibration \subref{figure:CalibResBottom}.}
	\label{figure:CalibRes}
\end{figure}

The calibration was performed on a hardware neuron, and the calibration data was stored in the database. 
Then, the biological parameters from the reference simulation were sent to the database, which provided the necessary hardware parameters in return. 
The floating gates of the corresponding neuron on the HICANN prototype were then programmed with these values. 
The results are shown in Figure~\ref{figure:CalibRes}. 
After calibration, the hardware neuron showed a firing rate of 536 kHz, which is very close to the reference simulation.

\section{Discussion}

Within the FACETS research project, a novel type of neuromorphic hardware system has been developed.
The device combines massive acceleration, large network sizes and a high configurability with the possible advantages inherent to analog designs, such as power efficiency and a time-continuous operation. 
Following this strategy, neuromorphic engineering has the potential to step out of its niche and provide new and relevant input to neuroscience, e.g.\ towards the understanding of cortical dynamics.
Still, as we noticed during the development of our specific system and during first experiments with prototypes, the originally available techniques and tools were clearly insufficient for exploiting the potential of such devices in a modeling context.
It is our experience that the quality of interfaces that make hardware flexibility actually usable is as essential as the electronic substrate itself.

The presented work approaches this challenge by introducing a methodological framework that establishes a balance between the configuration complexity and potential of a novel hardware system on the one hand and the usability and spectrum of possible applications for modelers without a hardware background on the other.
This neuromorphic modeling workflow has been depicted both in its conceptual whole and by means of detailed component descriptions.
It represents one major outcome of the inter-disciplinary collaboration among FACETS partners, thereby integrating expertise and progress in the fields of physiologically well-founded cortex modeling, hardware engineering and community-driven software development.
The multitude of the described components and their structured interaction reflects the comprehensiveness we are aiming at. 

We showed experimental data that provide a proof of mature functionality of the implemented stack of model-to-hardware translation tools.
The experimental results of mapping distortion studies on the basis of our virtual wafer-scale hardware system and reference software simulations represent examples of ongoing analysis work that continuously improves our software layer stack, the hardware design and our neuromorphic modeling experience.
A dedicated follow-up publication focusing on these analysis efforts is in preparation.
In particular, this work soon to be published will focus on computational aspects and address many questions that remain open at this point, especially concerning the computational and functional limitations that are imposed to the network models by the presented concepts.

The architectures in the presented benchmark collection already now cover a wide spectrum of computationally interesting aspects and relevant connectivity structures.
But although the workflow presented in this paper enables us to successfully realize these benchmarks with the FACETS hardware devices, further validation of the introduced concepts will be required. 
Important features like the synaptic plasticity mechanisms and large regions of the technically available hardware configuration space have not yet been systematically explored with our workflow in network contexts.
And, as soon as alternative neuromorphic devices with a comparable degree of configurability and size (but possibly different solutions and components) will be available from other groups, all applicable aspects of the described neuromorphic workflow will have to be tested also with these platforms.

In addition to such necessary investigations, we plan to extend the set of models that we use to benchmark and tune our workflow.
So far the realization of a large variety of biologically relevant structures has been the primary goal of iteratively applying the depicted optimization process.
A second focus will be put on computationally powerful architectures in general, independent of their biological plausibility.
Building upon this work, the presented methodological framework with the neuromorphic hardware system at its core will eventually be used to approach open neuroscientific questions.

\begin{acknowledgements}
The research leading to these results has received funding by the Sixth Framework Programme of the European Community (EC) under grant agreement no.\ 15879 (FACETS). 
Marc-Olivier Schwartz is supported by a doctoral fellowship in the Seventh Framework Programme of the EC under grant agreement no.\ 237955 (FACETS-ITN). 
Lyle Muller is supported by a doctoral fellowship from the \'Ecole des Neurosciences de Paris (ENP, Paris School of Neuroscience). 
\end{acknowledgements}

\bibliographystyle{spbasic}
\bibliography{references}
\addcontentsline{toc}{section}{References}

\end{document}